\documentclass[acmlarge]{acmart}

\AtBeginDocument{
  \providecommand\BibTeX{{
    \normalfont B\kern-0.5em{\scshape i\kern-0.25em b}\kern-0.8em\TeX}}}

\setcopyright{acmcopyright}
\copyrightyear{2020}
\acmYear{2020}
\acmDOI{10.1145/1122445.1122456}

\acmJournal{TOSEM}
\acmVolume{1}
\acmNumber{1}
\acmArticle{1}
\acmMonth{1}

\newcommand{\bl}[1]{\textcolor{black}{#1}}
\newcommand{\rd}[1]{\textcolor{red}{\textbf{#1}}}
\newcommand{\mc}[2]{\multicolumn{#1}{l}{#2}}

\usepackage{graphicx}
\usepackage{subfigure}
\graphicspath{figures/}
\DeclareGraphicsExtensions{.pdf,.jpeg,.png}

\usepackage{standalone}
\usepackage{longtable}
\usepackage{fancybox}
\usepackage{amsfonts}
\usepackage{amsmath}
\usepackage{fancyvrb}
\usepackage{color}
\usepackage{algorithmic}
\usepackage{framed}
\usepackage{array}
\usepackage{multirow}
\usepackage{booktabs}
\usepackage{balance}
\usepackage{flushend}

\PassOptionsToPackage{hyphens}{url}
\usepackage{hyperref}
\hypersetup{colorlinks=true,allcolors=blue}

\newcommand{\PreserveBackslash}[1]{\let\temp=\\#1\let\\=\temp}
\newcolumntype{C}[1]{>{\PreserveBackslash\centering}p{#1}}
\newcolumntype{R}[1]{>{\PreserveBackslash\raggedleft}p{#1}}
\newcolumntype{L}[1]{>{\PreserveBackslash\raggedright}p{#1}}

\definecolor{gray50}{gray}{.5}
\definecolor{gray40}{gray}{.6}
\definecolor{gray30}{gray}{.7}
\definecolor{gray20}{gray}{.8}
\definecolor{gray10}{gray}{.9}
\definecolor{gray05}{gray}{.95}

\newlength\Linewidth
\def\findlength{\setlength\Linewidth\linewidth
\addtolength\Linewidth{-4\fboxrule}
\addtolength\Linewidth{-3\fboxsep}
}
\newenvironment{examplebox}{\par\begingroup
  \setlength{\fboxsep}{5pt}\findlength
  \setbox0=\vbox\bgroup\noindent
  \hsize=0.95\linewidth
  \begin{minipage}{0.95\linewidth}\normalsize}
    {\end{minipage}\egroup
    \textcolor{gray20}{\fboxsep1.5pt\fbox
     {\fboxsep5pt\colorbox{gray05}{\normalcolor\box0}}}
    \endgroup\par\noindent
    \normalcolor\ignorespacesafterend}

\newcounter{RQCounter}

\newcommand{\RQ}[2]{%
\refstepcounter{RQCounter} \label{#1}
 \begin{center}	
  \begin{examplebox}
  \textbf{RQ\arabic{RQCounter}.}~#2
  \end{examplebox}	 
 \end{center}
}

\newcommand{\RS}[2]{%
\begin{framed}%
\filbreak
\noindent\textbf{Result {#1}:~}{\emph {#2}}%
\end{framed}
}

\newcommand{\SG}[2]{%
\begin{framed}%
\filbreak
\noindent\textbf{Suggestion {#1}:~}{\emph {#2}}%
\end{framed}
}

\newcommand{\OB}[2]{%
\begin{framed}%
\filbreak
\noindent\textbf{Observation {#1}:~}{\emph {#2}}%
\end{framed}
}

\begin{document}

\title{\bl{CodeMatcher: Searching Code Based on Sequential Semantics of Important Query Words}}

\author{Chao Liu}
\email{liuchaoo@zju.edu.cn}
\affiliation{
  \institution{College of Computer Science and Technology, Zhejiang University, China, and PengCheng Laboratory, China}}

\author{Xin Xia}
\email{Xia@monash.edu}

\authornote{Corresponding Author: Xin Xia.}
\affiliation{
  \institution{Monash University, Australia}}
  
\author{David Lo}
\email{davidlo@smu.edu.sg}
\affiliation{
 \institution{Singapore Management University, Singapore}}

\author{Zhiwei Liu}
\email{zhiweiliu03@baidu.com}
\affiliation{
 \institution{Baidu Inc., Shanghai, China.}}

\author{Ahmed E. Hassan}
\email{E-mail: ahmed@queensu.ca}
\affiliation{
  \institution{Queen's University, Ontario, Canada}}

\author{Shanping Li}
\email{shan@zju.edu.cn}
\affiliation{
  \institution{College of Computer Science and Technology, Zhejiang University, China}}

\renewcommand{\shortauthors}{Liu et al.}

\begin{abstract}

To accelerate software development, developers frequently search and reuse existing code snippets from a large-scale codebase, e.g., GitHub. Over the years, researchers proposed many information retrieval (IR) based models for code search, but they fail to connect the semantic gap between query and code. \bl{An early successful deep learning (DL) based model DeepCS solved this issue by learning the relationship between pairs of code methods and corresponding natural language descriptions. Two major advantages of DeepCS are the capability of understanding irrelevant/noisy keywords and capturing sequential relationships between words in query and code.} \bl{In this paper, we proposed an IR-based model CodeMatcher that inherits the advantages of DeepCS (i.e., the capability of understanding the sequential semantics in important query words), while it can leverage the indexing technique in the IR-based model to accelerate the search response time substantially. CodeMatcher first collects metadata for query words to identify irrelevant/noisy ones, then iteratively performs fuzzy search with important query words on the codebase that is indexed by the Elasticsearch tool, and finally reranks a set of returned candidate code according to how the tokens in the candidate code snippet sequentially matched the important words in a query.} We verified its effectiveness on a large-scale codebase with \textasciitilde{}41k repositories. Experimental results showed that CodeMatcher achieves an MRR (a widely used accuracy measure for code search) of \bl{0.60, outperforming DeepCS, CodeHow, and UNIF by 82\%, 62\%, and 46\% respectively.} Our proposed model is over \bl{1.2k} times faster than DeepCS. Moreover, CodeMatcher outperforms two existing online search engines (GitHub and Google search) by \bl{46\% and 33\%} respectively in terms of MRR. We also observed that: fusing the advantages of IR-based and DL-based models is promising; improving the quality of method naming helps code search, since method name plays an important role in connecting query and code.



\end{abstract}

\ccsdesc[500]{Software and its engineering~Search-based software engineering}

\keywords{code search, code indexing, mining software repositories, information retrieval}

\maketitle

\section{Introduction}\label{intro}
Code search is the most frequent activity in software development \cite{singer2010examination,sadowski2015developers,mcmillan2012recommending,shuaiimproving} as developers favor searching existing code and learning from them just-in-time when meeting a programming issue \cite{brandt2009two,brandt2010example}. Reusing existing diverse code from millions of open-source repositories (e.g., in GitHub \cite{kalliamvakou2014promises}) can maximize developers' productivity \cite{li2013help,sim2011well,stolee2014solving,mcmillan2011portfolio,gharehyazie2017some}. During software development, it was observed that more than 90\% of developers' code search efforts are used to find code snippet \cite{bajracharya2012analyzing}, thus this study focuses on searching code methods following previous studies \cite{gu2018deep,lv2015codehow,bajracharya2006sourcerer} instead of \bl{finding relevant repositories (e.g., from GitHub)} \cite{grechanik2010search,robillard2009recommendation}. \bl{Moreover, different from the studies of searching/synthesizing API usage examples \cite{buse2012synthesizing,holmes2009end}, our study focus on the query with natural language description (e.g., "how to convert InputStream to String") instead of the query with the expected API (class) names (e.g., "InputStream" or "StringBuilder"); the objective of our study is to directly find relevant code methods from codebase with similar semantics as the query, not to filter/synthesize representative API usage examples based on a set of candidate methods that were previously searched from codebase using the inputted API name.}

Prior works on code search studies start with leveraging information retrieval (IR) techniques (e.g., Koders \cite{bajracharya2012analyzing} and Krugle \cite{krugler2013krugle}), which regard method code as text and match keywords in query with indexed methods \cite{krugler2013krugle,bajracharya2006sourcerer}. But their performance is poor due to \bl{three} reasons: \textit{(1) short and diverse queries,} keywords matching can hardly represent various search requirements due to insufficient context; \textit{(2) representation of method as text,} method has a structure that carries specialized semantics \cite{bajracharya2012analyzing}\bl{; \textit{(3) poor ranking/sorting,} existing ranking approaches cannot precisely measure the semantic relevancy between query and candidate code    \cite{holmes2009end,keivanloo2014spotting}}. To address these issues, many past studies focus on query expansion and reformulation \cite{hill2011improving,haiduc2013automatic,lu2015query,mcmillan2011portfolio,ponzanelli2014mining}. For example, the Sourcerer model \cite{bajracharya2006sourcerer} extended the textual content of a method with structural information. The model proposed by Lu et al. \cite{lu2015query} expanded a query with synonyms generated from WordNet \cite{miller1998wordnet} and matched keywords to method signatures. The CodeHow model \cite{lv2015codehow} extended query with related APIs, and searched code methods with matched APIs and query keywords through an extended Boolean model\bl{. To improve the IR-based retrieval, Keivanloo et al. \cite{keivanloo2014spotting} clustered code methods based on the mined fine-grained code patterns. For each cluster, they selected one representative method to exclude duplicated methods with highly syntactic similarity. Afterward, they reranked the selected representative methods from all clusters in terms of code features (e.g., conciseness and completeness).}

Recently, Gu et al. \cite{gu2018deep} observed that IR-based models have two problems: \textit{(1) semantic gap,} keywords cannot adequately represent high-level intent implied in queries, and they also cannot reflect low-level implementation details in code; \textit{(2) representation gap,} query and code are semantically related, but they may not share common lexical tokens, synonyms, or language structures. To connect the semantic gap between query and code, \bl{early IR-based models mainly focused on expanding queries with related API (class) names \cite{zhang2017expanding,lv2015codehow}. To further solve the issue of representation gap,} Gu et al. \cite{gu2018deep} proposed a deep learning (DL) based model named DeepCS. It jointly embeds method code and natural language description into a high-dimensional vector space, where the methods with high similarities to a query are retrieved. Their experiments on a large-scale codebase collected from GitHub verified the model validity, showing substantial advantages over two representative IR-based models Sourcerer \cite{bajracharya2006sourcerer} and CodeHow \cite{lv2015codehow}.

Gu et al. \cite{gu2018deep} attributed the success of DeepCS, over IR-based models, to the capability of handling irrelevant/noisy keywords in queries and recognizing semantically related words between queries and code methods. However, DeepCS involves a complicated optimization process with a lot of manually set parameters. It requires more than 50 hours for model training on a machine with an Nvidia K40 GPU, and developers need to wait for about \bl{6 minutes} on a code search. Thus, it is necessary to explore the possibility to simplify DeepCS in a controllable way for practical usages. To address this challenge, we proposed an IR-based model CodeMatcher that maintains the advantageous features in DeepCS as described above. The IR technique is adopted because it can avoid time-consuming training and improve code search efficiency, and researchers do not need to take a lot of efforts for manually tuning complex parameters in DL-based model. Generally, CodeMatcher leverages Elasticsearch \cite{gormley2015elasticsearch}, a Lucene-based text search engine to index codebase and perform \bl{fuzzy search with the identified important keywords from search queries.} To improve query understanding, CodeMatcher removes noisy keywords according to some collected metadata and replaces irrelevant keywords with appropriate synonyms extracted from the codebase. To optimize the ranking of code methods searched by Elasticsearch, we designed two \bl{strategies} ($S_{name}$ and $S_{body}$) to rerank the code methods. For a pair of query and method, $S_{name}$ measures the semantic similarity between the query and method name. Meanwhile, $S_{body}$ measures the semantic similarity between the query and method body (i.e., the method implementation part). Like DeepCS, the similarity measurements consider the semantically related important words between queries and code methods and their sequential relationships. Our study investigated the following research questions (RQs):

\textit{\textbf{RQ1}: Can CodeMatcher outperform \bl{the baseline} models?} We tested CodeMatcher and the models DeepCS \cite{gu2018deep}, \bl{CodeHow \cite{lv2015codehow}, and UNIF \cite{cambronero2019deep}} on a large-scale testing data with \textasciitilde{}41k repositories. Experimental results showed that CodeMatcher achieves an MRR of 0.60, substantially outperforming DeepCS\bl{, CodeHow, and UNIF by 82\%, 62\%, and 46\%} respectively. 

\textit{\textbf{RQ2}: Is CodeMatcher faster than \bl{the baseline} models?} CodeMatcher needs no model training like DeepCS \bl{and UNIF}, and it only takes \bl{0.3s} for code search per query, which is \bl{over 1.2k and 1.5k times faster than two DL-based models DeepCS and UNIF respectively.} Besides, CodeMatcher \bl{works 8 times faster than} the IR-based model CodeHow.

\textit{\textbf{RQ3}: How do the CodeMatcher components contribute to the code search performance?} \bl{The CodeMatcher consists of three components, namely query understanding, fuzzy search, and reranking. The query understanding mainly collects metadata for the other two components. When removing the reranking component, the search performance is reduced by 22\% in terms of MRR (from 0.60 to 0.47). However, the model with only fuzzy search component still outperforms baseline models DeepCS, CodeHow, and UNIF by 42\%, 27\%, and 15\% respectively. Therefore, all the components are necessarily required for CodeMatcher.}


\textit{\textbf{RQ4}: Can CodeMatcher outperform existing online code search engines?} The proposed model CodeMatcher outperforms the popular online code search engines, GitHub search and Google search, by \bl{46\% and 33\%} in terms of MRR respectively. Moreover, we also found that combining Google search to CodeMatcher can further improve the MRR of CodeMatcher from \bl{0.60 to 0.64}. These results imply the merit of CodeMatcher in practical usage.

We offer these results to the software analytics community and suggest that sophisticated techniques like DL worth a try for innovation but \bl{their computation efficiency} should also be considered. Meanwhile, although the IR-based model performs better than DL-based models in terms of accuracy and time-efficiency in our experiment, the IR-based model cannot address queries with complex syntax and tolerate errors as the DL-based model. Thus, we recommend fusing the advantages of IR-based and DL-based models in the future. Besides, the high performance of CodeMatcher indicates that the method name is significant for code search, because it usually shares similar syntactic and semantic format as a search query. Thus, improving the quality of developers' method naming also helps code search.

The main contributions of this study are:

\begin{itemize}
    \item We propose an IR-based model CodeMatcher \bl{that not also runs fast but also inherits the advantages of} the DL-based model DeepCS. It can better understand query semantics by addressing irrelevant/noisy words, and correctly map the query-code semantics by capturing the sequential relationship between important words. \vspace{5pt}
    
    \item We evaluate the effectiveness of CodeMatcher on a large-scale codebase collected by us from GitHub. The experimental results show that CodeMatcher substantially outperforms \bl{three existing models (DeepCS, CodeHow, and UNIF)}.\vspace{5pt}
    
    \item We compare CodeMatcher with two existing online search engines, GitHub and Google search. CodeMatcher also shows substantial advantages over these two popular online search engines in code search.
\end{itemize}

The remainder of this paper is organized as follows. Section \ref{back} describes the background of the IR-based and DL-based models for code search. Section \ref{method} presents the proposed model CodeMatcher and shows how CodeMatcher is simplified from the DL-based model DeepCS. Section \ref{exp} and \ref{result} presents the experiment setup and results respectively, followed by the discussion in Section \ref{disucss} and the implication in Section \ref{implication}. Section \ref{related} describes related work, and Section \ref{threat} presents the threats to validation and model limitation. Finally, Section \ref{conclude} summarizes this study and presents future work.

\section{Background}\label{back}

As illustrated in Table \ref{tab_example}, for a search query (e.g., "how to read a text line by line in java"), code search models aim to find relevant code methods (e.g., "readTextLineByLine()") from a codebase with many candidate code methods. How to correctly map the semantics between the search query and code methods is the key to build a code search model. This section briefly introduces the background of \bl{two code search categories: IR-based and DL-based models.}

\begin{table}
    \centering
    \caption{Example of code search with a natural language query.}
    \begin{tabular}{|l|}
        \toprule
        \textbf{Query:} how to read a text line by line in java\\
        \midrule
        \bl{An irrelevant code:}\\
        public String \rd{HowToRead}(String path)\{\\
        \quad{}File\rd{Read}er fr = new File\rd{Read}er(path);\\
        \quad{}Buffered\rd{Read}er br = new Buffered\rd{Read}er(fr);\\
        \quad{}String first\rd{line} = br.\rd{readLine}();\\
        \quad{}return first\rd{line};\\
        \}\\
        \midrule
        \bl{A relevant code:}\\
        public void \rd{readTextLineByLine}(String file) \{\\
        \quad{}Buffered\rd{Read}er br = new Buffered\rd{Read}er(new FileInputStream(file));\\
        \quad{}String \rd{line} = null;\\
        \quad{}while ((\rd{line} = br.\rd{readLine}())!= null) \{\\
        \quad{}\quad{}System.out.println(\rd{line});\\
        \quad{}\}\\
        \quad{}br.close();\\
        \}\\
        \bottomrule
    \end{tabular}
    \label{tab_example}
\end{table}



\subsection{IR-Based Code Search}\label{back_ir}

Traditional code search models (e.g., CodeHow \cite{lv2015codehow}) mainly depend on IR techniques. Generally, the IR-based model builds an index for terms (i.e., each word token) in code methods, so that the keywords generated from search query can match the expected code methods quickly by checking the index. Afterward, the model recommends the top-n relevant code methods according to a designed keywords-terms matching degree. During the indexing, the method name and body are commonly regarded as two different components for a method \cite{lv2015codehow}. This is because method name (e.g., "readTextLineByLine()") is often defined in natural language whose semantic representation is close to the query (e.g., "how to read a text line by line"). But the method body implements the goal of the method name in programming languages (e.g., "new Buffered Reader()" and "br.close()"). Therefore, when matching keywords for a code method, a weighted sum of matching degrees on two different method components is utilized \cite{lv2015codehow}. Due to the success of existing search engines (e.g., Elasticsearch \cite{gormley2015elasticsearch}), code search researchers can easily build an IR-based model using the APIs provided by a search engine. The researchers can focus more on query understanding to address the semantic gap between query and code method. Namely, the IR-based models take more effort on how to generate correct keywords and how to calculate the keywords-terms matching degree. As illustrated in Table \ref{tab_example}, two code methods may contain many keywords related to the search query, but how to identify the relevant code and exclude the irrelevant one is still a challenging issue. CodeHow \cite{lv2015codehow} is an IR-based model. It leverages an Extended boolean model \cite{salton1983extended} to measure the degree to which query keywords match a method name and body respectively, and sort code methods based on a weighted sum of matching degrees for method name and body. \bl{To spot representative methods in the searched candidate methods, some models cluster candidates based on code patterns \cite{keivanloo2014spotting,mishne2012typestate,xie2006mapo}. For example, Keivanloo et al. \cite{keivanloo2014spotting} represented a method by an encoded pattern in terms of a set of code entities (e.g., method blocks). After retrieving a list of candidate methods for a query, they transformed them into vectors within a latent space where the methods with similar code patterns can be easily clustered. Afterward, they reranked representative methods in clusters based on their features (e.g., code conciseness and completeness).}

\subsection{DL-Based Code Search}\label{back_deepcs}

Gu et al. \cite{gu2018deep} indicated that existing IR-based models (e.g., CodeHow) do not work well because they have difficulties in query understanding. They cannot effectively address irrelevant/noisy keywords in the queries, so that the IR-based models cannot find code methods that highly related to a query \cite{gu2018deep}. To overcome this challenge, Gu et al. \cite{gu2018deep} provided a model DeepCS that leverages the DL technique to better understand query semantics and correctly search related code from a large scale of the codebase. Generally, to map the semantics of a search query and code methods, DeepCS aims to leverage DL the technique to vectorize queries and methods at first, and learn their semantic relationships directly. In this way, the methods relevant to a query can be easily identified by calculating the cosine similarities. In specific, the DL-based model DeepCS \cite{gu2018deep} encoded a pair of query and method into vectors with a fixed-size of vocabulary, and input the vectorized query-method pairs into different long-short term memory networks (LSTMs) \cite{hochreiter1997long}. The LSTM model aims to capture the sequential relationship between words in query or method. The parameters in the networks are optimized by a ranking loss function, which rewards the related query-method pairs and penalizes the unrelated ones. As the ground-truth query-method pair is not easy to obtain, DL-based models used the first line of method comment to represent the corresponding query. Similarly, the unrelated query to a method is the comment randomly selected from other commented methods. More details on DeepCS can be found in \cite{gu2018deep}.

\section{CodeMatcher}\label{method}
This section first presents the motivation of building a simplified model for DeepCS \cite{gu2018deep}, and then describe the implementation details of the proposed model CodeMatcher.

\subsection{Motivation.}

Different from the existing IR-based models, the DL-based model DeepCS can: (1) understand irrelevant/noisy keywords by word embedding technique, including synonyms and the words that have never appeared in the code methods; (2) capture sequential relationships between words in query and code methods via the LSTM model; and (3) map query intent to the semantics of code methods by measuring their semantic similarity \cite{gu2018deep}. However, due to the high complexity of DeepCS, it takes hours for model training and the code search is also time-consuming for practical usage. Therefore, it is necessary to find a way to simplify the model complexity but retain the beneficial features of the DL-based model. In this study, we simplify the DL-based model DeepCS into an IR-based model called CodeMatcher that maintains the advantages of DeepCS. Generally, we remove noisy keywords according to some collected metadata; replaces irrelevant keywords with appropriate synonyms extracted from codebase; design a code reranking approach by measuring the semantic similarity between query and code, which also considers the sequential relationship between words in query/code. The following subsection presents the implementation details of CodeMatcher.

\subsection{Implementation}\label{implementation}
\begin{figure}
    \centering
    \includegraphics[width=0.9\linewidth]{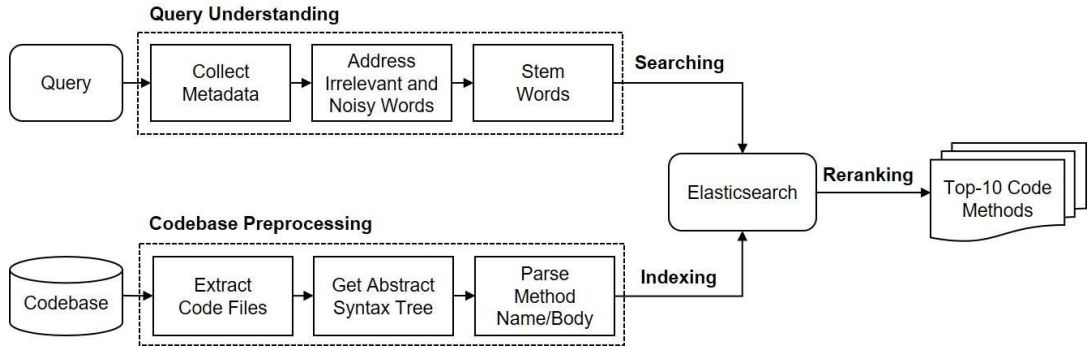}
    \caption{Overall framework of CodeMatcher.}
    \label{fig_workflow}
\end{figure}

The proposed model CodeMatcher performs code search in two phases following existing IR-based models, namely code indexing and code search, as described in Section \ref{back_ir}. Fig. \ref{fig_workflow} illustrates the overall framework of CodeMatcher. 

\vspace{5pt}\noindent\textbf{Phase-I: Codebase Preprocessing and Indexing.}
The first phase leverages the Elasticsearch \cite{gormley2015elasticsearch}, a Lucene-based text search engine for code indexing. Similar to existing IR-based models, we represent a code method as two components (i.e., method name and method body) and create an index for the method by calling the APIs provided by Elasticsearch. To extract method components from source code files, we developed a tool JAnalyzer\footnote{https://github.com/liuchaoss/janalyzer}, which transforms a method into an abstract syntax tree (AST) with the Javaparser\footnote{https://github.com/javaparser/javaparser} \cite{hosseini2013javaparser} library and extracts method name (e.g., "convertInputStreamToString()" in Table \ref{tab_example}) and body by traversing the AST. \bl{As JAnalyzer is depended on the Javaparser tool, thus it can only support the Java language instead of other programming languages.} Similar to the IR-based model CodeHow \cite{lv2015codehow}, the method body is parsed as a sequence of fully qualified tokens (e.g., Java.lang.String or System.io.File.readlines()) in method parameters, method body, and returned parameter.


\vspace{5pt}\noindent\textbf{Phase-II: Code Search.}
CodeMatcher takes three steps to perform code search on the indexed code methods for a query. The first step extracts some metadata for query understanding, and addresses noisy and irrelevant query words. The second step quickly finds a pool of highly related code methods from the codebase according to the collected metadata for query words. The last step reranks the searched methods in the second step according to the matching degrees between the query and candidate methods.


\underline{\textit{Step-1: Query understanding.}}
To address noisy words in a query, we removed the words that are not commonly used for programming. Specifically, we first identified word property (e.g., verb or noun) by using the Stanford Parser\footnote{https://nlp.stanford.edu/software/lex-parser.shtml} \cite{chen2014fast,zhang2017mining}. Then, we filtered out the question words and related auxiliary verb (e.g., how do), and excluded the verb-object/adpositional phrase on the programming language (e.g., in Java) as it is only used for the identification of programming language while our study only focuses on Java projects as Gu et al. \cite{gu2018deep}. Afterward, we removed the words that are not the verbs, nouns, adjectives, adverbs, prepositions, or conjunctions, since the removed words (e.g., 'a', 'the', and non-English symbols) are rarely used for coding \cite{arnold2000java}. Moreover, to identify the irrelevant words in a query, we counted the frequency of each word occurring in the method names of the codebase. We replaced the irrelevant words (i.e., the words not appeared in codebase) by their synonyms generated by the WordNet \cite{miller1998wordnet}. If the WordNet generates multiple synonyms for a word, we choose the synonym with the highest frequency in the codebase. Subsequently, we stemmed\footnote{https://pythonprogramming.net/stemming-nltk-tutorial/} the rest of query words to improve their generalizability in code search. 

Moreover, except for the words property (e.g., verb or noun) and frequency, we identified the third metadata named 'importance' for query words before performing code search. The word importance refers to how important a word used for programming. It can help CodeMatcher identify related important keywords between query and code methods. The importance is categorized by five levels as shown in Table \ref{tb_level}. For example, the most important one (i.e., level 5) is the JDK noun, e.g., "InputStream" or "readLine()". These metadata can help query understanding and forming keywords for code search. \bl{Verbs and nouns are important (level 4) as they can represent most of the semantics in code methods. This is because method/API names are typically made of verbs or verb phrases (such as "stop" or "sendMessage") while variable names are often written in nouns (e.g., "student" or "processor"), as described in the Google Java Style Guide\footnote{https://google.github.io/styleguide/javaguide.html}. We assigned JDK nouns with higher importance (level 5), because they are unique (namely it cannot be replaced by synonyms) and important to represent the semantics in code methods. For example, the JDK nouns "Inputstream" and "String" are important for the code with the semantics "convert Inputstream to String". Adjectives and adverbs are assigned lower importance (level 3) because they cannot indicate precise meaning without their corresponding noun/verb. Next, we assigned preposition/conjunction with the importance level 2. Finally, we assigned the other (often meaningless) symbols the lowest importance level (level 1).}

\begin{table}
    \centering
    \caption{Five word importance levels for programming based on the word property (e.g., verb or noun) and whether the word is a class name in JDK.}
    \begin{tabular}{|cll|}
    \toprule
        \textbf{Level} & \textbf{Condition} & \textbf{Examples} \\
    \midrule
        5 & JDK Noun & "Inputstream" and "readLine()"\\
        4 & Verb or Non-JDK Noun & "convert" and "whitespace"\\
        3 & Adjective or Adverb & "numeric" and "decimal"\\
        2 & Preposition or Conjunction & "from" and "or"\\
        1 & Other & Number and Non-English Symbols\\
    \bottomrule
    \end{tabular}
    \label{tb_level}
\end{table}

\underline{\textit{Step-2: Iterative fuzzy search.}}
To map the semantics of the preprocessed query and indexed code methods, we utilized the filtered query words in Step-1 to generate keywords for code search. Here we only match the keywords with indexed method names to quickly narrow the search space. To perform code search, we launched an iterative fuzzy match on indexed method names with a regular string\footnote{https://docs.python.org/3/library/re.html}. The string is formed by all remaining query words in order as "$.*word_{1}.*\cdots.*word_{n}.*$". We used the regular match instead of the term match to capture the sequential relationship between words as DeepCS \cite{gu2018deep}. For example, "convert int to string" should be different from "convert string to int". Afterward, if the total number of returned methods is no more than ten, we removed the least important word with lower frequency one at a time according to their metadata, and performed the fuzzy search again until only no query word left. For each search round, we filtered out the repeated method by comparing the MD5 hash\footnote{https://docs.python.org/2/library/hashlib.html} values of their source code.

\underline{\textit{Step-3: Reranking.}}
To refine the method rankings returned from Step-2, we designed a metric ($S_{name}$) to measure the matching degree between query and method name as Eq. (\ref{eq_score}). A larger value of $S_{name}$ indicates a higher-ranked method with more overlapped tokens between query and method name in order. Moreover, if equal $S_{name}$ exists, we boosted the rank of a method with a higher matching score ($S_{body}$) between query and method body, as calculated in Eq. (\ref{eq_score2}). Similar to $S_{name}$, a higher $S_{body}$ value implies better token matching between query and tokens in method body orderly. However, different from $S_{name}$, we added the last term in $S_{body}$ to represent the ratio of JDK APIs in method, in terms of the fully qualified tokens, because developers favor a method with more basic APIs (e.g., JDK) \cite{sadowski2015developers,martie2015sameness}. After the above two rounds of ranking refinement, we returned the top-10 methods in the list. 

\bl{Note that, although our tool JAnalyzer can extract Javadocs and comments for code, the CodeMatcher did not consider them when reranking the searched code following all the related studies \cite{gu2018deep,lv2015codehow, cambronero2019deep,feng2020codebert,husain2019codesearchnet}. This is because the code search task aims to solve the semantic gap between the query in natural language and the code in programming language. Besides, whether the comments and Javadocs written in natural language are helpful for code search is highly dependent on their quality and number. However, in practical usage, we cannot assume that any code always contains high-quality comment or Javadoc.}

\begin{equation}\label{eq_score}
\begin{split}
    S_{name} &= \frac{\#query\ words\ as\ keywords}{\#query\ words} \\
    &\times{} \frac{\#characters\ in\ name\ orderly\ matched\ keywords}{\# characters\ in\ name}\\
\end{split}
\end{equation}

\begin{equation}\label{eq_score2}
\begin{split}
    S_{body} &= \frac{\#API\ words\ matched\ query\ words}{\#query\ words} \\
    &\times{} \frac{Max[\#API\ words\ orderly\ matched\ query\ words]}{\#query\ words}\\
    &\times{} \frac{\#JDK APIs}{\#APIs}\\
\end{split}
\end{equation}

\vspace{5pt}\noindent\textbf{Example.}
Fig. \ref{codematcher_example} illustrated an example for the first query \textit{"convert an inputstream to a string"} in Table \ref{queries50}. From the token metadata, we can notice that both 'inputstream' and 'string' have level-5 importance (i.e., they are JDK objects), and they are frequently used for method naming (frequency$>$3442). With this metadata, CodeMatcher successively generates four candidate regular match strings on indexed method names. For the two returned methods, the first one ranked higher due to its larger matching scores on method name ($S_{name}$) and body ($S_{body}$). 

\begin{table}
    \centering
    \caption{An example for CodeMatcher}
    \begin{tabular}{|l|}
    \begin{tabular}{|l|c|c|c|c|c|c|}
        \toprule
         \mc{7}{\textbf{Query:} convert an inputstream to a string} \\
         \midrule
         \mc{7}{\textbf{(0) Indexed Codebase}} \\
         \midrule
         \mc{7}{\bl{Source Code 1:}}\\
         \mc{7}{ public String \textbf{convertInputStreamToString}(InputStream is)\{}\\
         \mc{7}{\quad{}InputStreamReader isr = new InputStreamReader(is);}\\
         \mc{7}{\quad{}BufferedReader r = new BufferedReader(isr);}\\
         \mc{7}{StringBuilder sb = new StringBuilder();}\\
         \mc{7}{\quad{}String line;}\\
         \mc{7}{\quad{}while ((line = r.readLine()) != null)\{}\\
         \mc{7}{\quad{}\quad{}sb.append(line);}\\
         \mc{7}{\quad{}\}}\\
         \mc{7}{\quad{}return sb.toString();}\\
         \mc{7}{\}}\\
         \mc{7}{\bl{Method Name: }\rd{convertInputStreamToString}}\\
         \mc{7}{\bl{API Sequence: }java.io.\rd{InputStream}, java.io.\rd{InputStream}Reader, java.lang.\rd{String}Builder,}\\
         \mc{7}{java.lang.\rd{String}, java.lang.\rd{String}Builder.readline(), java.lang.\rd{String}Builder.append(),}\\
         \mc{7}{java.lang.\rd{String}Builder.\rd{toString}(), java.io.\rd{String}}\\
         \midrule
         \mc{7}{\bl{Source Code 2:}
         public String convertInputStream2String(InputStream is)\{}\\
         \mc{7}{\quad{}return convert(is);}\\
         \mc{7}{\}}\\
         \mc{7}{\bl{Method Name: }\rd{convertInputStream}2\rd{String}}\\
         \mc{7}{\bl{API Sequence: }java.io.\rd{InputStream}, Util.\rd{convert}(), java.lang.\rd{String}}\\
         \midrule
         \mc{7}{\textbf{(1) Token Metadata}}\\
         \midrule
         \bl{Token} & convert & an & inputstream & to & a & string\\
         \midrule
         \bl{Property} & verb & other & noun & prep & other & noun \\
         \midrule
         \bl{Frequency} & 39292 & 0 & 3442 & 22 & 0 & 52369\\
         \midrule
         \bl{Importance} & 4 & 1 & 5 & 2 & 1 & 5\\
         \midrule
         \mc{7}{\textbf{(2) Keywords for Code Search}}\\
         \midrule
         \mc{7}{\bl{Regular Match String 1:} .*convert.*inputstream.*to.*string.*}\\
         \mc{7}{\bl{Regular Match String 2:} .*convert.*inputstream.*string.*}\\
         \mc{7}{\bl{Regular Match String 3:} .*inputStream.*string.*}\\
         \mc{7}{\bl{Regular Match String 4:} .*String.*}\\
         \midrule
         \mc{7}{\textbf{(3) Reranking}}\\
         \midrule
         \mc{7}{\bl{Rank = 1, }\textbf{convertInputStreamToString()\{...\}}}\\
         \mc{7}{\rd{$S_{name}=\frac{4}{6}\times\frac{26}{26}=0.67$}, \rd{$S_{body}=\frac{3}{6}\times\frac{3}{6}\times\frac{8}{8}=0.25$}}\\
         \midrule
         \mc{7}{\bl{Rank = 2, }\textbf{convertInputStream2String()\{...\}}}\\
         \mc{7}{\rd{$S_{name}=\frac{3}{6}\times\frac{24}{25}=0.48$}, \rd{$S_{body}=\frac{3}{6}\times\frac{2}{6}\times\frac{2}{3}=0.11$}}\\
        \bottomrule
    \end{tabular}
    \end{tabular}
    \label{codematcher_example}
\end{table}


\section{Experiment Setup}\label{exp}
This section describes the investigated research questions, the collected large-scale dataset for model validation, and the widely used model evaluation criteria.

\subsection{Research Questions}\label{rqs}

To verify the validity of the proposed model, this study investigates the following research questions (RQs):

\RQ{1}{Can CodeMatcher outperform \bl{the baseline} models?}

The proposed model CodeMatcher aims to leverage IR technique to simplify the complexity of DeepCS while retaining its advantages, namely its capability of addressing irrelevant/noisy keywords in queries and recognizing semantically related words between queries and code methods as described by Gu et al. \cite{gu2018deep}. To verify the validity of the proposed model, this RQ investigates whether the CodeMatcher outperforms the relevant models, \bl{including two DL-based model DeepCS \cite{gu2018deep} and UNIF \cite{cambronero2019deep}, and an IR-based model CodeHow \cite{lv2015codehow}.}


\RQ{2}{Is CodeMatcher faster than \bl{the baseline} models?}

We observed that training and testing the DL-based model DeepCS is time-consuming due to its high complexity. Therefore, we intend to analyze whether the simplified model CodeMatcher works faster than DeepCS substantially. Besides, we also compare the time-efficiency of CodeMatcher with \bl{UNIF \cite{cambronero2019deep} and CodeHow \cite{lv2015codehow}}.

\RQ{3}{\bl{How do the CodeMatcher components contribute to the code search performance?}}

CodeMatcher consists of three important components: \bl{the fuzzy search component retrieves an initial set of candidate relevant code from the indexed codebase, the reranking component sorts the candidate list based on the designed strategy, and the query understanding component collects metadata for the previous two components.} They largely determine the performance of the code search. Thus, this RQ aims to analyze how much these components contribute to the model performance. The result can also help analyze the necessity of these components.


\RQ{4}{Can CodeMatcher outperform existing online code search engines?}

GitHub search\footnote{https://github.com/search} and Google search\footnote{https://google.com} are two commonly used search engines for developers to find code in practice. To measure the performance of GitHub/Google search, all the search engines are tested with the same queries as CodeMatcher and they are set to search only GitHub repositories; for Google search, we use the following advanced settings: "site:github.com" and "filetype:java".






\subsection{Dataset}\label{data}

\vspace{5pt}\noindent\textbf{Codebase.}
Originally, Gu et al. \cite{gu2018deep} collected 9,950 Java projects that have at least 20 stars in GitHub as the testing data of DeepCS. However, we cannot use their testing data to verify our IR-based model CodeMatcher, because Gu et al. \cite{gu2018deep} just provided a preprocessed data that can be used for DeepCS only. Therefore, we built a new and larger scale testing data for model evaluation. We crawled 41,025 Java repositories from GitHub created from Jul. 2016 to Dec. 2018 with more than five stars. The number of stars filter ($>$5 stars) which is different from Gu et al.'s \cite{gu2018deep} setting (i.e., at least 20 stars) is used so that the testing data includes more Java projects. Besides, the time duration (Jul. 2016 to Dec. 2018) of our new testing data can ensure the non-overlapping with DeepCS' training data (created from Aug. 2008 to Jun. 2016). Table \ref{codebase} shows that the new codebase contains \textasciitilde{}17 million methods, and 21.91\% of them have Javadoc comments that describe the corresponding methods.

\begin{table}
    \centering
    \caption{Statistics of the constructed codebase.}
    \begin{tabular}{|cccc|}
        \toprule
        \textbf{$\#$Project} & \textbf{$\#$Method} & \textbf{$\#$Javadoc} & \textbf{\#LOC} \\
        \midrule
        41,025 & 16,611,025 & 3,639,794 & 70,332,245\\
        \bottomrule
    \end{tabular}
    \label{codebase}
\end{table}

\vspace{5pt}\noindent\textbf{Queries.}
To simulate a real-world code search scenario, we validated a code search model \bl{with three query sets in total of 174 queries} as listed in \bl{Tables \ref{queries50}-\ref{queries25}}:

\begin{itemize}
    \item \textit{$Queries_{50}$.} \bl{The first query set was} manually collected by Gu et al. \cite{gu2018deep} from Stack Overflow in a systematic way. These queries are top-50 voted Java programming questions\footnote{https://stackoverflow.com/questions/tagged/java?sort=votes} following three criteria: \textit{(1) concrete,} a question should be a specific programming task, such as "How can I concatenate two arrays in Java?"; \textit{(2) well-answered,} the accepted answers corresponding to the question should contain at least one code snippet; \textit{(3) non-duplicated,} the question is not a duplicate of another question in the same collection.\vspace{5pt}
    
    \item \textit{$Queries_{99}$.} \bl{The second query set was collected from the CodeSearchNet challenge\footnote{https://github.com/github/CodeSearchNet} that was built by Husain et al. \cite{husain2019codesearchnet}. The contained 99 queries\footnote{https://github.com/github/CodeSearchNet/blob/master/resources/queries.csv} were common search queries from Bing that have high clickthrough rates to code with clearly technical keywords \cite{husain2019codesearchnet}.}\vspace{5pt}
    
    \item \textit{$Queries_{25}$.} \bl{The third query set has 25 queries based on the studies of Mishne et al. \cite{mishne2012typestate} and Keivanloo et al. \cite{keivanloo2014spotting}. Note that these two studies used API names as code search input. However, our study focused on the queries written in natural language. Therefore, we used the natural language descriptions provided in these two studies \cite{mishne2012typestate,keivanloo2014spotting} as our model inputs.}
\end{itemize}

\begin{table}[b]
\scriptsize
\caption{\bl{The Queries$_{50}$ collected from Gu et al. \cite{gu2018deep}} and the Evaluation Results (NF: Not Found within the top 10 returned results, DCS: DeepCS \cite{gu2018deep}, CM: CodeMatcher, CH: CodeHow \cite{lv2015codehow}, UNIF \cite{cambronero2019deep}.)}
\begin{tabular}{|c|l|cccc|}
\toprule
\textbf{No.} & \textbf{Query}     & \textbf{DCS} & \textbf{CM} & \textbf{CH} & \textbf{UNIF} \\  \midrule
1  & convert an inputstream to a string     & 2  & 1 & 2 & 2   \\ 
2  & create arraylist from array  & 8  & 1 & 1 & 4   \\ 
3  & iterate through a hashmap    & 1  & 1 & 1 & NF  \\ 
4  & generating random integers in a specific range   & 1  & 1 & 5 & 4   \\ 
5  & converting string to int in java & 8  & 1 & 2 & 1   \\ 
6  & initialization of an array in one line & 1  & 1 & 1 & 2   \\ 
7  & how can I test if an array contains a certain value & NF & 1 & 1 & 9   \\ 
8  & lookup enum by string value  & NF & 1 & 1 & NF  \\ 
9  & breaking out of nested loops in java   & NF & 3 & NF   & NF  \\ 
10 & how to declare an array      & 1  & 1 & 1 & NF  \\ 
11 & how to generate a random alpha-numeric string    & 1  & 1 & NF   & 5   \\ 
12 & what is the simplest way to print a java array   & 1  & 1 & NF   & 5   \\ 
13 & sort a map by values  & 5  & 1 & 1 & 1   \\ 
14 & fastest way to determine if an integer’s square root is an   integer & NF & NF   & NF   & NF  \\ 
15 & how can I concatenate two arrays in java  & 9  & 1 & NF   & 3   \\ 
16 & how do I create a java string from the contents of a file  & 3  & NF   & 5 & 2   \\ 
17 & how can I convert a stack trace to a string      & 2  & 1 & 1 & 2   \\ 
18 & how do I compare strings in java & NF & 1 & 1 & NF  \\ 
19 & how to split a string in java   & 1  & 1 & 10   & 9   \\ 
20 & how to create a file and write to a file in java & NF & 3 & 4 & 8   \\ 
21 & how can I initialise a static map      & 2  & 2 & 1 & 4   \\ 
22 & iterating through a collection, avoiding concurrent   modification exception when removing in loop & 5  & NF   & 10   & 2   \\ 
23 & how can I generate an md5 hash  & 4  & 1 & 1 & 1   \\ 
24 & get current stack trace in java & 1  & 1 & 1 & 1   \\ 
25 & sort arraylist of custom objects by property     & 2  & NF   & 7 & 8   \\ 
26 & how to round a number to n decimal places in java   & 1  & 1 & 3 & 2   \\ 
27 & how can I pad an integers with zeros on the left & 8  & NF   & 10   & 2   \\ 
28 & how to create a generic array in java  & 3  & 1 & NF   & 2   \\ 
29 & reading a plain text file in java      & 4  & 3 & 3 & 1   \\ 
30 & a for loop to iterate over enum in java   & NF & 1 & NF   & NF  \\ 
31 & check if at least two out of three booleans are true & NF & NF   & NF   & NF  \\ 
32 & how do I convert from int to string    & 10 & 1 & 4 & 10  \\ 
33 & how to convert a char to a string in java & 6  & 1 & NF   & 1   \\ 
34 & how do I check if a file exists in java   & NF & 1 & 3 & 1   \\ 
35 & java string to date conversion  & NF & 1 & 1 & NF  \\ 
36 & convert inputstream to byte array in java & 1  & 1 & 4 & 1   \\ 
37 & how to check if a string is numeric in java      & 2  & 1 & NF   & 1   \\ 
38 & how do I copy an object in java & NF & 5 & 7 & NF  \\ 
39 & how do I time a method's execution in java & NF & 1 & NF   & 4   \\ 
40 & how to read a large text file line by line using java      & 8  & NF   & NF   & 1   \\ 
41 & how to make a new list in java  & 4  & 1 & 1 & NF  \\ 
42 & how to append text to an existing file in java   & 1  & NF   & NF   & NF  \\ 
43 & converting iso 8601-compliant string to date     & 9  & NF   & 5 & 2   \\ 
44 & what is the best way to filter a java collection & NF & 2 & NF   & NF  \\ 
45 & removing whitespace from strings in java  & NF & 1 & NF   & 1   \\ 
46 & how do I split a string with any whitespace chars as   delimiters    & NF & NF   & 1 & 2   \\ 
47 & in java, what is the best way to determine the size of an   objects  & NF & 1 & NF   & NF  \\ 
48 & how do I invoke a java method when given the method name as a   string  & NF & NF   & NF   & NF  \\ 
49 & how do I get a platform dependent new line character & NF & NF   & NF   & NF  \\ 
50 & how to convert a map to list in java   & 7  & 1 & NF   & 4   \\ \bottomrule
\end{tabular}
\label{queries50}
\end{table}

\begin{table}[]
\scriptsize
\caption{\bl{The Queries$_{99}$ collected from Husain et al. \cite{husain2019codesearchnet}} and the Evaluation Results (NF: Not Found within the top 10 returned results, DCS: DeepCS \cite{gu2018deep}, CM: CodeMatcher, CH: CodeHow \cite{lv2015codehow}, UNIF \cite{cambronero2019deep}.)}
\begin{tabular}{|c|L{80pt}|cccc||c|L{120pt}|cccc|}
\toprule
\textbf{No.} & \textbf{Query} & \textbf{DCS} & \textbf{CM} & \textbf{CH} & \textbf{UNIF} & 
\textbf{No.} & \textbf{Query} & \textbf{DCS} & \textbf{CM} & \textbf{CH} & \textbf{UNIF} \\ \midrule
1 & convert int to string & 2 & 1 & 6  & 2   & 51   & how to randomly pick a number    & 6 & 5 & NF   & 1 \\ 
2  & priority queue & NF & NF   & NF   & 1   & 52   & normal distribution    & NF   & 1 & NF   & NF   \\ 
3  & string to date & 9  & 1 & 1 & 6   & 53   & nelder mead optimize   & NF   & NF   & NF   & NF   \\ 
4  & sort string list  & 1  & 1 & NF   & 7   & 54   & hash set for counting distinct   elements  & 7 & NF   & NF   & NF   \\ 
5  & save list to file & 2  & 1 & NF   & NF  & 55   & how to get database table name   & NF   & 2 & NF   & NF   \\ 
6  & postgresql connection    & NF & 1 & 3 & NF  & 56   & deserialize json & 1 & 1 & 1 & 1 \\ 
7  & confusion matrix  & NF & 6 & NF   & 2   & 57   & find int in string     & 5 & 1 & NF   & 1 \\ 
8  & set working directory    & NF & 1 & NF   & 1   & 58   & get current process id & 6 & 1 & NF   & 1 \\ 
9  & group by count & NF & 2 & 2 & NF  & 59   & regex case insensitive & NF   & 2 & NF   & 10   \\ 
10 & binomial distribution    & NF & 7 & NF   & 1   & 60   & custom http error response & 7 & 1 & 2 & NF   \\ 
11 & aes encryption & 5  & 1 & 3 & 7   & 61   & how to determine a string is a valid   word   & NF   & NF   & NF   & NF   \\ 
12 & linear regression & NF & 2 & 4 & 1   & 62   & html entities replace  & NF   & NF   & 1 & 1 \\ 
13 & socket recv timeout      & NF & 1 & 5 & 1   & 63   & set file attrib hidden & NF   & NF   & NF   & NF   \\ 
14 & write csv      & 1  & 2 & 6 & 1   & 64   & sorting multiple arrays based on   another arrays sorted order & 5 & NF   & 7 & NF   \\ 
15 & convert decimal to hex   & 1  & 1 & NF   & 1   & 65   & string similarity levenshtein    & NF   & 2 & 6 & 1 \\ 
16 & export to excel   & NF & 1 & 3 & 4   & 66   & how to get html of website & NF   & 5 & 3 & NF   \\ 
17 & scatter plot   & NF & 1 & 4 & NF  & 67   & buffered file reader read text   & 1 & 1 & 1 & NF   \\ 
18 & convert json to csv      & NF & NF   & 4 & NF  & 68   & encrypt aes ctr mode   & NF   & NF   & 1 & NF   \\ 
19 & pretty print json & 1  & 1 & NF   & 1   & 69   & matrix multiply & NF   & 1 & 1 & 1 \\ 
20 & replace in file   & 1  & 1 & NF   & 2   & 70   & print model summary    & NF   & NF   & NF   & NF   \\ 
21 & k means clustering & NF & 3 & 1 & 3   & 71   & unique elements & 1 & 1 & NF   & 1 \\ 
22 & connect to sql & 1  & 1 & 1 & 1   & 72   & extract data from html content   & 4 & NF   & 2 & 1 \\ 
23 & html encode string & 1  & 2 & 4 & 1   & 73   & heatmap from 3d coordinates      & NF   & NF   & 3 & NF   \\ 
24 & finding time elapsed using a timer & NF & 1 & NF   & 5   & 74   & get all parents of xml node      & NF   & 3 & 1 & NF   \\ 
25 & parse binary file to custom class  & NF & NF   & NF   & NF  & 75   & how to extract zip file recursively & NF   & 1 & 7 & 4 \\ 
26 & get current ip address   & 2  & 1 & NF   & 1   & 76   & underline text in label widget   & NF   & NF   & NF   & NF   \\ 
27 & convert int to bool      & 2  & 1 & NF   & 8   & 77   & unzipping large files  & NF   & 1 & 2 & NF   \\ 
28 & read text file line by line & 5  & NF   & 1 & 1   & 78   & copying a file to a path  & NF   & 1 & 1 & NF   \\ 
29 & get executable path      & 1  & 8 & 1 & 1   & 79   & get the description of a http status   code   & 5 & 1 & 4 & 1 \\ 
30 & httpclient post json     & 2  & 3 & 1 & NF  & 80   & randomly extract x items from a list & 1 & NF   & NF   & NF   \\ 
31 & get inner html & 4  & 3 & 2 & 2   & 81   & convert a date string into yyyymmdd & 1 & 1 & 4 & 2 \\ 
32 & convert string to number & 8  & 1 & 1 & 2   & 82   & convert a utc time to epoch      & NF   & NF   & NF   & NF   \\ 
33 & format date    & 1  & 1 & 1 & 1   & 83   & all permutations of a list & NF   & 1 & NF   & 1 \\ 
34 & readonly array & NF & NF   & NF   & NF  & 84   & extract latitude and longitude from   given input    & 5 & NF   & NF   & 2 \\ 
35 & filter array   & 2  & 1 & 1 & 1   & 85   & how to check if a checkbox is checked      & NF   & 2 & NF   & 2 \\ 
36 & map to json    & 2  & 3 & 2 & 3   & 86   & converting uint8 array to image  & 1 & NF   & NF   & NF   \\ 
37 & parse json file   & NF & 1 & NF   & 2   & 87   & memoize to disk  - persistent memoization  & NF   & NF   & NF   & NF   \\ 
38 & get current observable value & NF & NF   & NF   & NF  & 88   & parse command line argument      & 1 & 1 & 3 & 1 \\ 
39 & get name of enumerated value & NF & NF   & NF   & 1   & 89   & how to read the contents of a .gz   compressed file? & NF   & NF   & 1 & NF   \\ 
40 & encode url     & 1  & 1 & 4 & 3   & 90   & sending binary data over a serial   connection & NF   & NF   & NF   & NF   \\ 
41 & create cookie  & 1  & 1 & NF   & NF  & 91   & extracting data from a text file & 6 & NF   & 9 & 1 \\ 
42 & how to empty array & NF & NF   & NF   & NF  & 92   & positions of substrings in string   & 1 & 1 & 1 & 6 \\ 
43 & how to get current date  & NF & 2 & 1 & NF  & 93   & reading element from html -   \textless{}td\textgreater{}      & NF   & 1 & NF   & NF   \\ 
44 & how to make the checkbox checked   & NF & 1 & 2 & NF  & 94   & deducting the median from each column      & NF   & NF   & 1 & NF   \\ 
45 & initializing array & 2  & 1 & 10   & 4   & 95   & concatenate several file remove   header lines & NF   & 1 & NF   & NF   \\ 
46 & how to reverse a string  & NF & 1 & 1 & 5   & 96   & parse query string in url & 2 & NF   & 9 & 2 \\ 
47 & read properties file     & 1  & 1 & NF   & 1   & 97   & fuzzy match ranking    & NF   & NF   & NF   & NF   \\ 
48 & copy to clipboard & 1  & 1 & 1 & 3   & 98   & output to html file    & NF   & NF   & 2 & NF   \\ 
49 & convert html to pdf      & NF & 1 & 2 & NF  & 99   & how to read .csv file in an efficient   way   & NF   & NF   & NF   & NF   \\ 
50 & json to xml conversion   & NF & 1 & NF   & NF  & \multicolumn{1}{l|}{} &    & \multicolumn{4}{l|}{} \\ \bottomrule
\end{tabular}
\label{queries99}
\end{table}

\begin{table}
\scriptsize
\caption{\bl{The Queries$_{25}$ collected from Mishne et al. \cite{mishne2012typestate} and Keivanloo et al. \cite{keivanloo2014spotting}}, and the Evaluation Results (NF: Not Found within the top 10 returned results, DCS: DeepCS \cite{gu2018deep}, CM: CodeMatcher, CH: CodeHow \cite{lv2015codehow}, UNIF \cite{cambronero2019deep}.)}
\begin{tabular}{|c|l|cccc|}
\toprule
\textbf{No.} & \textbf{Query} & \textbf{DCS} & \textbf{CM} & \textbf{CH} & \textbf{UNIF} \\ \midrule
1  & upload a file  & 1  & 3 & 5 & 1   \\ 
2  & parse a command-line and get values   from it   & NF & 1 & 1 & NF  \\ 
3  & prepare the executable and argument   of a command-line   & NF & 2 & NF   & NF  \\ 
4  & create a path element and append it   to existing and boot paths & NF & NF   & NF   & NF  \\ 
5  & run a query and iterate over the   results   & NF & NF   & NF   & NF  \\ 
6  & commit and then rollback a   transaction     & NF & 2 & 1 & 1   \\ 
7  & get the key of an array element type  & 1  & NF   & 6 & NF  \\ 
8  & get the description and nature IDs of   a Java project & NF & NF   & NF   & NF  \\ 
9  & create a new action      & 2  & 1 & 1 & 6   \\ 
10 & get the input for the current editor  & NF & 2 & NF   & NF  \\ 
11 & retrieve arguments from command line  & NF & 1 & 1 & 6   \\ 
12 & check user selection     & 6  & 1 & NF   & 1   \\ 
13 & set up a ScrollingGraphicalViewer  & NF & NF   & NF   & NF  \\ 
14 & create a project  & 1  & 2 & NF   & 3   \\ 
15 & successfully login and logout      & NF & 1 & NF   & NF  \\ 
16 & click an Element  & 1  & 1 & 1 & 2   \\ 
17 & commit and rollback a statement    & NF & NF   & NF   & 3   \\ 
18 & send a HTTP request via URLConnection & 6  & 1 & 2 & 1   \\ 
19 & redirect Runtime exec() output with   System & NF & NF   & NF   & NF  \\ 
20 & get OS Level information such as   memory    & NF & NF   & NF   & 1   \\ 
21 & SSH Connection & NF & 5 & NF   & 3   \\ 
22 & download and save a file from network & NF & 1 & NF   & 1   \\ 
23 & generate a string-based MD5 hash   value     & NF & NF   & 6 & 2   \\ 
24 & read the content of a HttpResponse   object line by line  & NF & NF   & NF   & NF  \\ 
25 & search via Lucene and manipulate the   hits  & NF & NF   & 1 & NF  \\ \bottomrule
\end{tabular}
\label{queries25}
\end{table}

\subsection{Baseline Models and Replication Package.}\label{baselines}

In the experiment, \bl{three} models are selected as our baseline models, including Gu et al.'s \cite{gu2018deep} DeepCS, Lv et al.'s \cite{lv2015codehow} CodeHow\bl{, and Cambronero et al.'s UNIF \cite{cambronero2019deep}}. To test baseline models on our codebase, we first preprocessed Java code files within all projects by our tool JAnalyzer\footnote{https://github.com/liuchaoss/janalyzer}. It first parses the abstract syntax tree of each Java code file by leveraging the Javaparser\footnote{https://github.com/javaparser/javaparser} library, and extracts all necessary method components as the inputs of baselines, such as method name, comment, Javadoc, APIs in the method body, etc. Note that we do not test these models on DeepCS' original testing data but our new data, because DeepCS provides no raw data (i.e., no source code) but preprocessed data that can be only used by DeepCS. To mitigate the replication difficulty for our study, we provide a replication package\footnote{\textbf{Replication Package: https://bitbucket.org/ChaoLiuCQ/codematcher}} to share our codebase, and source code of CodeMatcher and baseline models. The compared baseline models are described as follows.

\vspace{5pt}\noindent\textbf{DeepCS,} 
the DL-based model proposed by Gu et al. \cite{gu2018deep}: We trained DeepCS by re-running the source code\footnote{https://github.com/guxd/deep-code-search} and pre-processed training data\footnote{https://pan.baidu.com/s/1U$\_$MtFXqq0C-Qh8WUFAWGvg} provided by the authors. To test DeepCS on our codebase, we first did some natural language processing (e.g., stemming) following the description in Gu et al.'s paper \cite{gu2018deep}, then encoded data using DeepCS' vocabulary, and saved the encoded data into the required format by using DeepCS' internal APIs in source code. 



\vspace{5pt}\noindent\textbf{CodeHow,}
the IR-based model developed by Lv et al. \cite{lv2015codehow}: It expands a query with words in related official APIs and matches it with code methods. As Lv et al. \cite{lv2015codehow} provide no source code and data for replication, we re-implemented CodeHow strictly following their paper. Our implementation can be found in our shared replication package. Note that as CodeHow was used for searching C\# projects while our targets are Java repositories, we thus used the JDK (Java development kit) as the source of official APIs.

\vspace{5pt}\noindent\textbf{\bl{UNIF,}}
\bl{the DL-based model proposed by Cambronero et al. \cite{cambronero2019deep}: Similar to DeepCS, the UNIF transforms code and query into vectors and it is trained by pairs of code and natural language description. However, the main differences are that: UNIF represents code by method name and a bag of tokens, which is a subset of DeepCS' input; the UNIF used the pre-trained fastText \cite{bojanowski2017enriching} as its embedding layer. Details can be found in \cite{cambronero2019deep}. As the author provided no source code, we re-implemented it by ourselves strictly following the description in the original study, which is also provided in our replication package.}




\subsection{Evaluation Criteria}\label{criteria}
To measure the effectiveness of code search models, we need to identify the relevancy of a returned code method to a query. Following Gu et al. \cite{gu2018deep}, the relevancy is manually identified by two independent developers, and the disagreements are resolved by open discussions. During the relevance identification, developers only consider the top-10 returned code methods. Based on the identified relevancy, we measure model performance using four widely used evaluation metrics following Gu et al. \cite{gu2018deep}, including FRank, SuccessRate@k, Precision@k, and Mean Reciprocal Rank (MRR). Note that FRank is defined only for one query while the other metrics are defined on all queries.

\vspace{5pt}\noindent\textbf{FRank}, is the rank of the first correct result in the result list \cite{gu2018deep}. It measures users' inspection efforts for finding a relevant method when scanning the candidate list from top to bottom. A smaller FRank value implies fewer efforts and better effectiveness of a code search tool for a particular query.

\vspace{5pt}\noindent\textbf{SuccessRate}@k, the percentage of queries for which more than one correct result exists in the top-k ranked results \cite{gu2018deep}. Specifically, $SuccessRate@k=Q^{-1}\sum_{q=1}^{Q}\delta(FRank_{q}\leq{k})$, where $Q$ is the total number of tested queries; $\delta(\cdot)$ is an indicator function that returns 1 if the input is true and 0 otherwise. Higher SuccessRate@k means better code search performance, and users can find desired method by inspecting fewer returned method list.

\vspace{5pt}\noindent\textbf{Precision@k}, is the average percentage of relevant results in top-k returned method list for all queries. It is calculated by $Precision@k=Q^{-1}\sum_{q=1}^{Q}r_{q}/k$, where $Q$ is the total number of queries; $r_{q}$ is the number of related results for a query $q$ \cite{gu2018deep}. Precision@k is useful and important because users often check many returned results for learning different code usages \cite{raghothaman2016swim}. Larger Precision@k indicates that a code search model returns less noisy results.

\vspace{5pt}\noindent\textbf{MRR}, the average of the reciprocal ranks for all queries, where the reciprocal rank of a query is the inverse of the rank of the first relevant result ($FRank$) \cite{gu2018deep}. Thus, the formula is $MRR=Q^{-1}\sum_{q=1}^{Q}FRank_{q}^{-1}$. Larger MRR value means a higher ranking for the first relevant methods.


\section{Results}\label{result}
This section provides the experimental results to the four investigated RQs raised in Section \ref{rqs}.

\subsection{Can CodeMatcher Outperform the \bl{Baseline} Models?}\label{rq1}

\begin{table}[]
\caption{Performance comparison of CodeMatcher and baseline models (DeepCS, CodeHow, and UNIF), where the model performance is measured by SuccessRate@1/5/10 (SR@1/5/10), Precision@1/5/10 (P@1/5/10), and MRR.}
\begin{tabular}{|c|l|ccc|ccc|c|}
\toprule
\textbf{Queries} & \textbf{Model} & \textbf{SR@1} & \textbf{SR@5} & \textbf{SR@10} & \textbf{P@1}  & \textbf{P@5}  & \textbf{P@10} & \textbf{MRR}  \\ \midrule 
\multirow{4}{*}{$Queries_{50}$}   & DeepCS      & 0.22 & 0.46 & 0.64  & 0.22 & 0.23 & 0.22 & 0.36 \\ 
                                & CodeHow     & 0.30 & 0.52 & 0.62  & 0.30 & 0.23 & 0.21 & 0.41 \\  
                                & UNIF        & 0.22 & 0.58 & 0.68  & 0.22 & 0.21 & 0.17 & 0.40 \\ 
                                & CodeMatcher & 0.64 & 0.76 & 0.76  & 0.64 & 0.58 & 0.57 & 0.71 \\ \midrule
\multirow{4}{*}{$Queries_{99}$}   & DeepCS      & 0.20 & 0.38 & 0.45  & 0.20 & 0.14 & 0.11 & 0.33 \\ 
                                & CodeHow     & 0.22 & 0.45 & 0.54  & 0.22 & 0.21 & 0.19 & 0.36 \\ 
                                & UNIF        & 0.33 & 0.49 & 0.56  & 0.33 & 0.21 & 0.15 & 0.43 \\  
                                & CodeMatcher & 0.48 & 0.65 & 0.68  & 0.48 & 0.42 & 0.37 & 0.58 \\ \midrule
\multirow{4}{*}{$Query_{25}$}   & DeepCS      & 0.16 & 0.20 & 0.28  & 0.16 & 0.06 & 0.06 & 0.26 \\ 
                                & CodeHow     & 0.24 & 0.32 & 0.40  & 0.24 & 0.15 & 0.14 & 0.34 \\ 
                                & UNIF        & 0.24 & 0.44 & 0.52  & 0.24 & 0.22 & 0.15 & 0.38 \\ 
                                & CodeMatcher & 0.28 & 0.56 & 0.56  & 0.28 & 0.26 & 0.19 & 0.46 \\ \midrule
\multirow{4}{*}{$Queries_{all}$}  & DeepCS      & 0.20 & 0.38 & 0.48  & 0.20 & 0.15 & 0.13 & 0.33 \\ 
                                & CodeHow     & 0.25 & 0.45 & 0.54  & 0.25 & 0.20 & 0.19 & 0.37 \\  
                                & UNIF        & 0.29 & 0.51 & 0.59  & 0.29 & 0.21 & 0.16 & 0.41 \\ 
                                & CodeMatcher & 0.50 & 0.67 & 0.68  & 0.50 & 0.44 & 0.40 & 0.60 \\ \bottomrule
\end{tabular}
\label{tab:results}
\end{table}

\begin{figure}
    \centering
    \includegraphics[width=0.9\linewidth]{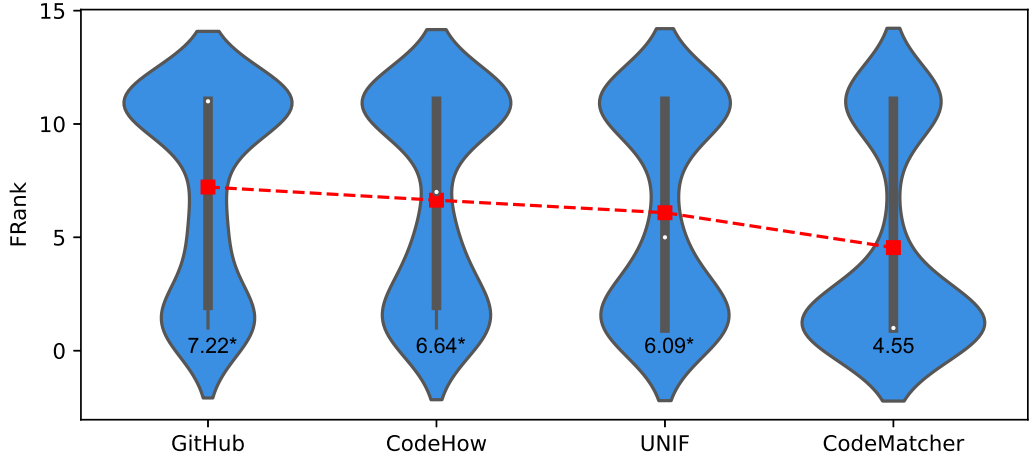}
    \caption{Violin plots of FRank for CodeMatcher and baseline models (DeepCS, CodeHow, and UNIF), where red squares and white dots indicate the mean and median FRank respectively; the values denote the mean FRank for each model; '*' indicates the significant difference (p-value $<$ 0.05) between a model and CodeMatcher, which is tested by the Wilcoxon signed-rank test \cite{wilcoxon1945individual} at a 5\% significance level.}
    \label{fig_frank_baseline}
\end{figure}

Table \ref{tab:results} compares the experimental results between the proposed model CodeMatcher and \bl{three baseline models (i.e., DeepCS, CodeHow, and UNIF)} on our large-scale testing data. We can notice that, for the studied 174 queries ($Queries_{all}$), DeepCS obtains an \bl{MRR of 0.33, where SuccessRate@1/5/10 = 0.20/0.38/0.48 and Precision@1/5/10 = 0.20/0.15/0.13.} Meanwhile, CodeHow achieves an \bl{MRR of 0.37 with SuccessRate@1/5/10 = 0.25/0.45/0.54 and Precision@1/5/10 = 0.25/0.20/0.19.} Besides, \bl{UNIF gains a better search performance with MRR = 0.41, where SuccessRate@1/5/10 = 0.29/0.51/0.59 and Precision@1/5/10 = 0.29/0.21/0.16.}


Table \ref{tab:results} shows that the proposed model CodeMatcher achieves an \bl{MRR of 0.60 with SuccessRate@1/5/10 = 0.50/0.67/0.68 and Precision@1/5/10 = 0.50/0.44/0.40.} Comparing with the baseline models \bl{(DeepCS, Codehow, and UNIF)}, the MRR improved by \bl{81.8\%, 62.2\%, and 46.3\%} respectively; SuccessRate@1/5/10 increased by \bl{150\%/76.3\%/41.7\%, 100\%/48.9\%/25.9\%, and 72.4\%/31.4\%/15.3\%} respectively; Precision@1/5/10 boosted by \bl{150\%/193.3\%/207.7\%, 100\%/120\%/110.5\%, and 72.4\%/109.5\%/150\%} respectively. Moreover, Fig. \ref{fig_frank_baseline} shows the violin plots of FRank between models, where CodeMatcher obtains better mean FRank (4.55) over the other baselines ($>$6). CodeMatcher's advantage in FRank is also statistically significant (p-value$<$0.05), after we tested the FRank between CodeMatcher and a model with the Wilcoxon signed-rank test \cite{wilcoxon1945individual} at a 5\% significance level. The above experimental results imply that the CodeMatcher performs well and clearly outperforms existing solutions.

\RS{1}{Our CodeMatcher outperforms the baseline models \bl{(DeepCS, CodeHow, and UNIF)} substantially, indicating the simplification from DeepCS to CodeMatcher is reasonable and valuable.}


\subsection{Is CodeMatcher Faster than \bl{the Baseline} Models?}\label{codematcher_evaluation}

Table \ref{tb_time} compares the time efficiency between CodeMatcher and baseline models \bl{(DeepCS, CodeHow, and UNIF)}. All these models ran on a server with an Intel-i7 CPU and an Nvidia Geforce GTX1060 GPU. \bl{In the practical usage, developers expect that code search models can quickly respond their search query, namely short searching time. For the IR-based models (CodeMatcher and CodeHow), they can quickly retrieve a small subset of methods from codebase by leveraging the indexing technique (commonly with <1s for a query) and sort the candidates with a light-weight ranking strategy. Note that the indexing technique needs to build indexing for all methods in advance to facilitate the code search. Although building indexing may take a long time, it is acceptable as building indexing commonly works offline. Meanwhile, to return the top-n relevant methods for a query, the DL-based models (DeepCS and UNIF) have to compute the cosine similarities between query and all methods among codebase in terms of high-dimension vectors. As DL-based models cannot accelerate the search time like the IR-based models with the indexing technique, they have to leverage the multi-threading technique to boost their performance.} 

The results show that DeepCS took 58.16h to train, 24.51h to preprocess codebase (i.e., parse, encode, and vectorize method name/APIs/tokens), and 376.4s to search code for each query. \bl{In contrast, UNIF only spent 4.1h for model training because its network is simple which contains only embedding and attention layers instead of the complex LSTM layer used in DeepCS. However, UNIF required 455.3s to complete the searching task for a query, much slower than DeepCS, because UNIF represented code/query into vector with 500 dimensions \cite{cambronero2019deep} while DeepCS only needed vectors with 400 dimensions \cite{gu2018deep}. Although the dimension just increased by 25\%, UNIF takes 21\% more time for code search as the codebase is large with more than 16 million code as described in Table \ref{codebase}. Therefore, the similarity computation for the higher dimensional vectors takes much more time, even if the UNIF has a simple network.}

Comparing with DeepCS \bl{and UNIF}, our IR-based model CodeMatcher ran faster and \bl{substantially decreased the code search time from 370+s to 0.3s per query.} We can notice that CodeMatcher did not need a long-time training \bl{as DL-based models, although the training usually happens rarely (or just once).} As to the searching time, CodeMatcher only required 23.5h to preprocess code (23.2h for code parsing and 0.3h for code indexing). \bl{We found that the DL-based models work slowly mainly because it is time-consuming ($>$300s for a query) to load the 23+GB vectorized codebase (with 16m methods) for computing their 
cosine similarities with the query even using the multi-threading technique. But the IR-based models do not have this issue, because they leveraged the indexing technique to quickly ($<$1s for a query) reduce the search space scale (from 16m methods to hundreds of methods or less) so that the code search can be done in a very short time. Therefore, further DL-based studies need to consider ways to solve this bottleneck.} Additionally, CodeMatcher works \bl{8 times faster than} the IR-based model CodeHow mainly because \bl{the fuzzy search component in CodeMatcher only retrieved a limited number of candidate code from codebase as described in Section \ref{implementation}.}

\begin{table}
    \centering
    \caption{Time efficiency comparison between CodeMatcher and baseline models in three different phases.}
    \setlength{\tabcolsep}{12pt}{
    \begin{tabular}{|l|rrr|}
        \toprule
        \textbf{Model}        & \textbf{Train} & \textbf{Preprocess} & \textbf{Search}\\
        \midrule
        DeepCS      & 58.2h & 24.5h & 376.4s/query\\
        UNIF        & 4.1h & 24.2h & 455.3s/query \\
        CodeHow    & - & 23.5h & 2.4s/query\\
        CodeMatcher  & -      & 23.5h & 0.3s/query \\
        \bottomrule
    \end{tabular}}
    \label{tb_time}
\end{table}

The 23.5 hours of code preprocessing seems time-consuming for CodeMatcher. However, there are about 17 million methods, as shown in Table \ref{tb_time}, and each method only takes about 0.005s for code preprocessing on average. \bl{The low preprocessing time for the IR-based models means that the model can quickly parse and index the update or added methods in codebase.} Thus, CodeMatcher can support the dynamic and rapidly expanding code scale of GitHub as this model requires no optimization, where changed or new methods can be rapidly parsed and indexed. 


\RS{2}{Our model CodeMatcher is faster than the DL-based models DeepCS \bl{and UNIF}, because CodeMatcher requires no model training and it can process a query with \bl{more than 1.2k times} speedup. Meanwhile, the CodeMatcher works \bl{8 times faster than} the IR-based model CodeHow.}




\subsection{\bl{How do the CodeMatcher Components Contribute to the Code Search Performance?}}\label{why_well}

\begin{figure}
    \centering
    \includegraphics[width=0.9\linewidth]{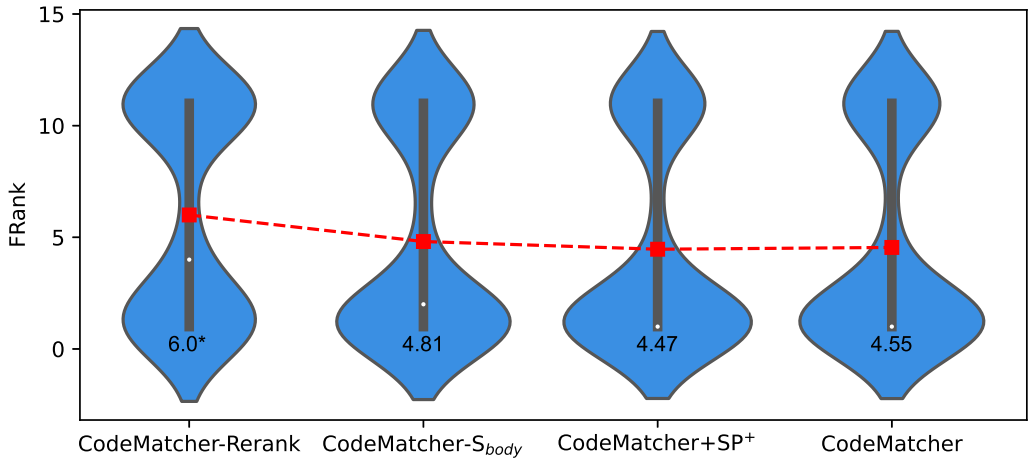}
    \caption{Violin plots of FRank for CodeMatcher \bl{with different settings (-Rerank, excluding the reranking component; -S$_{body}$, excluding the ranking strategy $S_{body}$ in the reranking component; +SP$^{+}$, using the corrected results generated by the Stanford Parser)}, where red squares and white dots indicate the mean and median FRank respectively; the values denote the mean FRank for each model; '*' indicates the significant difference (p-value $<$ 0.05) between a model and CodeMatcher, which is tested by the Wilcoxon signed-rank test \cite{wilcoxon1945individual} at a 5\% significance level.}
    \label{fig_frank_component}
\end{figure}

\bl{Generally, the CodeMatcher consists of three components: query understanding, fuzzy search, and reranking, As described in Section \ref{implementation}. The query understanding aims to collect metadata for fuzzy search and reranking. Note that the query understanding component leveraged the Stanford Parser to identify the property of query words (e.g., noun and verb). However, the property cannot be precisely identified due to the limitation of the Stanford Parser. We found that the parser failed to work for 124 words in 7 cases, which affects 55.2\% of the 174 queries. To investigate how the parser's accuracy affects the model performance. We manually corrected the wrong cases, and applied them to the code search (CodeMatcher+SP$^{+}$). Table \ref{tab:components} shows that CodeMatcher+SP$^{+}$ obtains an MRR of 0.61, outperforming CodeMatcher by only 2\%. Fig. \ref{fig_frank_component} also indicates that the mean FRank (4.47 vs. 4.55) between CodeMatcher+SP$^{+}$ and CodeMatcher is not significantly different (p-value$>$0.05) after performing the Wilcoxon signed-rank test \cite{wilcoxon1945individual} at a 5\% significance level. We found that the optimization seldom improved the search performance because of two reasons. One is that some failed identifications do not influence the importance level for words as described in Table \ref{tb_level}, including "Noun->Verb" and "Verb->Noun". The other reason is that many queries are short with a limited number of words so that the other two components can still find the expected code. Thus, under this situation, the accuracy of the used Stanford Parser is acceptable for CodeMatcher.}

\bl{Moreover, to investigate how the key components (fuzzy search and reranking) affect the model performance, we tested the CodeMatcher by excluding the reranking (CodeMatcher-Rerank). Table \ref{tab:components} shows that the MRR of CodeMatcher-Rerank is 0.47, which is reduced by 21.7\% comparing with the MRR of CodeMatcher. Besides, Fig. \ref{fig_frank_component} shows that the mean FRank value is increased from 4.55 to 6.0, where the difference is significant (p-value$<$0.05). These results indicate that the reranking plays an important role in searching code. Besides, we can observe that even though the fuzzy search component outperforms the other baseline models (DeepCS, CodeHow, and UNIF) in terms of MRR (<0.42) by 42.4\%, 27.0\%, and 14.7\% respectively.}


\begin{table}[]
\caption{Performance comparison of CodeMatcher with different component settings (\bl{-Rerank, excluding the reranking component; -S$_{body}$, excluding the ranking strategy $S_{body}$ in the reranking component; +SP$^{+}$, using the corrected results generated by the Stanford Parser}), where the model performance is measured by SuccessRate@1/5/10 (SR@1/5/10), Precision@1/5/10 (P@1/5/10), and MRR.}
\begin{tabular}{|c|l|ccc|ccc|c|}
\toprule
\textbf{Queries}               & \textbf{Model}          & \textbf{SR@1} & \textbf{SR@5} & \textbf{SR@10} & \textbf{P@1} & \textbf{P@5} & \textbf{P@10} & \textbf{MRR} \\ \midrule
\multirow{4}{*}{$Queries_{50}$}  & CodeMatcher             & 0.64          & 0.76          & 0.76           & 0.64         & 0.58         & 0.57          & 0.71         \\ 
                               & CodeMatcher-Rerank      & 0.54          & 0.68          & 0.68           & 0.54         & 0.42         & 0.40          & 0.63         \\  
                               & CodeMatcher-$S_{body}$ & 0.60          & 0.72          & 0.74           & 0.60         & 0.53         & 0.50          & 0.68         \\  
                               & CodeMatcher+$SP^+$    & 0.66          & 0.76          & 0.76           & 0.66         & 0.59         & 0.58          & 0.72         \\ \midrule
\multirow{4}{*}{$Queries_{99}$}  & CodeMatcher             & 0.48          & 0.65          & 0.68           & 0.48         & 0.42         & 0.37          & 0.58         \\
                               & CodeMatcher-Rerank      & 0.33          & 0.51          & 0.58           & 0.33         & 0.29         & 0.25          & 0.45         \\ 
                               & CodeMatcher-$S_{body}$ & 0.45          & 0.62          & 0.67           & 0.45         & 0.37         & 0.33          & 0.56         \\  
                               & CodeMatcher+$SP^+$    & 0.49          & 0.66          & 0.69           & 0.49         & 0.42         & 0.38          & 0.60         \\ \midrule
\multirow{4}{*}{$Queries_{25}$}  & CodeMatcher             & 0.28          & 0.56          & 0.56           & 0.28         & 0.26         & 0.19          & 0.46         \\  
                               & CodeMatcher-Rerank      & 0.12          & 0.16          & 0.24           & 0.12         & 0.11         & 0.09          & 0.21         \\ 
                               & CodeMatcher-$S_{body}$ & 0.32          & 0.52          & 0.52           & 0.32         & 0.28         & 0.22          & 0.46         \\ 
                               & CodeMatcher+$SP^+$    & 0.32          & 0.56          & 0.56           & 0.32         & 0.30         & 0.26          & 0.48         \\ \midrule
\multirow{4}{*}{$Queries_{all}$} & CodeMatcher             & 0.50          & 0.67          & 0.68           & 0.50         & 0.44         & 0.40          & 0.60         \\  
                               & CodeMatcher-Rerank      & 0.36          & 0.51          & 0.56           & 0.36         & 0.30         & 0.27          & 0.47         \\ 
                               & CodeMatcher-$S_{body}$ & 0.48          & 0.63          & 0.67           & 0.48         & 0.40         & 0.37          & 0.58         \\ 
                               & CodeMatcher+$SP^+$    & 0.52          & 0.67          & 0.69           & 0.52         & 0.45         & 0.42          & 0.61         \\ \bottomrule
\end{tabular}
\label{tab:components}
\end{table}

\begin{table}[]
\caption{\bl{Wrong cases that the Stanford Parser in CodeMatcher identified the property of query words.}}
\begin{tabular}{|c|c||c|c|}
\toprule
\textbf{Wrong Case}         & \textbf{No.} & \textbf{Wrong Case} & \textbf{No.} \\ \midrule
Noun-\textgreater{}Verb       & 67 & Noun-\textgreater{}Adjective  & 7          \\ 
Verb-\textgreater{}Noun       & 23 & Verb-\textgreater{}Adjective  & 2         \\ 
Adjective-\textgreater{}Verb  & 18 & Conjective-\textgreater{}Noun & 1         \\ 
Adjective-\textgreater{}Noun  & 6  & \textbf{Total} & 124          \\ \bottomrule
\end{tabular}
\label{wrong_cases_for_sp}
\end{table}

For CodeMatcher described in Section \ref{method}, $S_{name}$ and $S_{body}$ are two matching scores to determine the ranks of searched results, but it is unknown how much they attributed to the performance of CodeMatcher. Thus, we used CodeMatcher for code search but removing the matching score $S_{body}$. Table \ref{tab:results} shows that the model only using $S_{name}$ (CodeMatcher-$S_{body}$) obtains \bl{MRR = 0.47}. Comparing with standard CodeMatcher, MRR is reduced by \bl{21.7\%}. Besides, the violin plots in Fig.  \ref{fig_frank_component} shows that the mean FRank \bl{(4.55 vs. 4.81)} between CodeMatcher and CodeMatcher-$S_{body}$ are not significantly different (p-value $>$ 0.05 for the Wilcoxon signed-rank test \cite{wilcoxon1945individual} at a 5\% significance level). These results indicate that the score $S_{name}$ that matches query keywords with method name dominated the performance of CodeMatcher, and implies that method name is a significant bridge for the semantic gap between query and code. Furthermore, although the influence of $S_{body}$ that matches query keywords with method body is low, we cannot ignore its contribution. We observed that $S_{body}$ did not work as well as $S_{name}$ because it cannot fully connect the semantics of query and method body, which are written in natural language and programming language respectively. Therefore, the $S_{body}$ part requires further improvement.

\RS{3}{\bl{All the CodeMatcher components are necessarily required.} CodeMatcher works well mainly because it can precisely match the query with relevant methods in names by considering the importance of the programming words in the query and the order of query tokens.}

\subsection{Can CodeMatcher Outperform Existing Online Code Search Engines?}\label{rq4}


\begin{table}[]
\caption{Performance comparison of CodeMatcher and online search engines (Google and GitHub search), where the model performance is measured by SuccessRate@1/5/10 (SR@1/5/10), Precision@1/5/10 (P@1/5/10), and MRR.}
\begin{tabular}{|c|l|ccc|ccc|c|}
\toprule
\textbf{Queries}               & \textbf{Model}     & \textbf{SR@1} & \textbf{SR@5} & \textbf{SR@10} & \textbf{P@1} & \textbf{P@5} & \textbf{P@10} & \textbf{MRR} \\ \midrule
\multirow{4}{*}{$Queries_{50}$}  & CodeMatcher        & 0.64          & 0.76          & 0.76           & 0.64         & 0.58         & 0.57          & 0.71         \\ 
                               & GitHub Search      & 0.28          & 0.60          & 0.64           & 0.28         & 0.21         & 0.17          & 0.44         \\ 
                               & Google Search      & 0.32          & 0.80          & 0.90           & 0.32         & 0.34         & 0.30          & 0.51         \\ 
                               & CodeMatcher+Google & 0.72          & 0.90          & 0.94           & 0.72         & 0.42         & 0.34          & 0.80         \\ \midrule
\multirow{4}{*}{$Queries_{99}$}  & CodeMatcher        & 0.48          & 0.65          & 0.68           & 0.48         & 0.42         & 0.37          & 0.58         \\ 
                               & GitHub Search      & 0.27          & 0.57          & 0.64           & 0.27         & 0.21         & 0.17          & 0.42         \\ 
                               & Google Search      & 0.27          & 0.56          & 0.66           & 0.27         & 0.29         & 0.24          & 0.42         \\
                               & CodeMatcher+Google & 0.49          & 0.71          & 0.76           & 0.49         & 0.34         & 0.26          & 0.61         \\ \midrule
\multirow{4}{*}{$Queries_{25}$}  & CodeMatcher        & 0.28          & 0.56          & 0.56           & 0.28         & 0.26         & 0.19          & 0.46         \\ 
                               & GitHub Search      & 0.16          & 0.40          & 0.48           & 0.16         & 0.14         & 0.14          & 0.31         \\ 
                               & Google Search      & 0.28          & 0.52          & 0.60           & 0.28         & 0.24         & 0.24          & 0.41         \\  
                               & CodeMatcher+Google & 0.32          & 0.56          & 0.68           & 0.32         & 0.25         & 0.24          & 0.45         \\ \midrule
\multirow{4}{*}{$Queries_{all}$} & CodeMatcher        & 0.50          & 0.67          & 0.68           & 0.50         & 0.44         & 0.40          & 0.60         \\ 
                               & GitHub Search      & 0.26          & 0.55          & 0.61           & 0.26         & 0.20         & 0.17          & 0.41         \\ 
                               & Google Search      & 0.29          & 0.62          & 0.72           & 0.29         & 0.30         & 0.26          & 0.45         \\
                               & CodeMatcher+Google & 0.53          & 0.74          & 0.80           & 0.53         & 0.35         & 0.28          & 0.64         \\ \bottomrule
\end{tabular}
\label{tab:search_engine}
\end{table}

\begin{figure}
    \centering
    \includegraphics[width=0.9\linewidth]{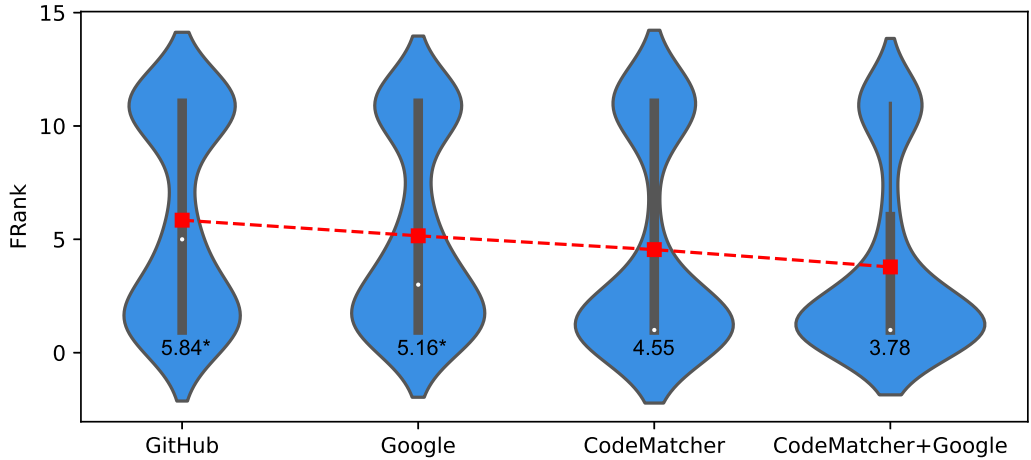}
    \caption{Violin plots of FRank for CodeMatcher and compared models (GitHub search, Google search, CodeMatcher + Google), where red squares and white dots indicate the mean and median FRank respectively; the values denote the mean FRank for each model; '*' indicates the significant difference (p-value $<$ 0.05) between a model and CodeMatcher, which is tested by the Wilcoxon signed-rank test \cite{wilcoxon1945individual}.}
    \label{fig_frank_engine} 
\end{figure}

GitHub\footnote{https://github.com/search} and Google\footnote{https://www.google.com/} search engines are what developers commonly used for code search in the real world. Comparing CodeMatcher with GitHub/Google search is helpful for understanding the usefulness of CodeMatcher. To investigate the advantages of CodeMatcher over the GitHub/Google search, we used the \bl{174 queries in Tables \ref{queries50}-\ref{queries25}} as their inputs and performed code search on the whole Java code methods in GitHub. To work on GitHub codebase, the Google search engine performs code search with the following advanced settings: "site:github.com" and "filetype:java". 

However, we need to note that GitHub and Google search engines are different from CodeMatcher in three aspects: \textit{(1) Larger-scale codebase.} the codebase of GitHub and Google contains more repositories, because these two online search engines cannot control the search scope as our collected codebase for CodeMatcher. \textit{(2) Wider context.} GitHub/Google search matches keywords in a query on all text in code files (e.g., method, comment, and Javadoc) while CodeMatcher only uses texts on methods. \textit{(3) Code snippet vs. method.} GitHub/Google search returns a list of code snippets, but they would not necessarily be a method as what CodeMatcher returns. When using the GitHub and Google search engines for code search, both of them return a list of code snippets with matched keywords, and then we inspect top-10 code snippets to check whether the snippets are relevant to the corresponding queries. We excluded the snippet that was returned just because its comment records the Stack Overflow link (e.g., \textit{"https://stackoverflow.com/questions/157944/create-arraylist-from-array"}) with the same string as the search query (e.g., "create arraylist from array"). This is because the query is also collected from the question title of that link, so that this kind of code snippet will overestimate the performance of GitHub/Google search. \bl{As CodeMatcher searches code not based on comment or Javadoc, we do not need to filter the Stack Overflow links as GitHub and Google searchers.}

Table \ref{tab:search_engine} shows that the GitHub search achieves \bl{MRR = 0.41 with SuccessRate@1/5/10 = 0.26/0.55/0.61 and Precision@1/5/10 = 0.26/0.20/0.17.} Meanwhile the Google search obtains an \bl{MRR of 0.45, where SuccessRate@1/5/10 = 0.29/0.62/0.72 while Precision@1/5/10 = 0.29/0.30/0.26} respectively. We can notice that CodeMatcher outperforms GitHub and Google search engines by \bl{46.3\% and 33.3\%} respectively in terms of MRR; by \bl{92.3\%/21.8\%/11.5\% and 72.4\%/8.1\%/-5.6\%} in terms of SuccessRate@1/5/10; by \bl{92.3\%/120\%/135.3\% and 72.4\%/46.7\%/53.8\%} in terms of Precision@1/5/10. Additionally, as illustrated in Fig. \ref{fig_frank_engine}, CodeMatcher achieves better mean FRank \bl{(4.55)} over GitHub/Google search (mean FRank $>$ 5.1); the difference is statistically significant (p-value$<$0.05 tested by the Wilcoxon signed-rank test \cite{wilcoxon1945individual} at a 5\% significance level). These results indicate that CodeMatcher shows advantages in SuccessRate@1/5, Precision@1/5/10, and MRR as compared with the existing two online search engines. These experimental results indicate the practical merit of the CodeMatcher over the GitHub/Google search.

Moreover, we can notice that although Google cannot achieve high precision as CodeMatcher if returning only one method for a query, it can successfully recommend at least one relevant method for more queries. Due to this observation, we investigate the code search performance of combining CodeMatcher and Google together, where the first method returned by Google is replaced by the one recommended by CodeMatcher. The last row in Table \ref{tab:search_engine} shows that the combined model (CodeMatcher+Google) gains the best SuccessRate@1/5/10 \bl{(0.53/0.74/0.80), Precision@1 (0.53), and MRR (0.64)} compared with other models in the Table, even if there are some sacrifices in Precision@5/10 \bl{(0.35/0.28)}. \bl{We found that CodeMatcher+Google gains higher SuccessRate@1 than CodeMatcher because CodeMatcher cannot find any results for some queries which can be compensated by Google's results. Specifically, the CodeMatcher and Google cannot return any correct code for 31.6\% and 28.2\% of total queries respectively when we inspected the top-10 returned results. But the adopted combination strategy (CodeMatcher+Google) can reduce the failure rate to 20\%.} We can also notice that the combined model (CodeMatcher+Google) improved the mean FRanks of CodeMatcher \bl{(4.55) and Google (5.16) to 3.78} as illustrated in Fig. \ref{fig_frank_engine}, although the improvement over CodeMatcher is not statistically significant (p-value$>$0.05 tested by the Wilcoxon signed-rank test \cite{wilcoxon1945individual} at a 5\% significance level). This case implies that it is beneficial to incorporate the proposed model CodeMatcher into the Google search engine. 

\RS{4}{CodeMatcher is also advantageous over existing online search engines, GitHub and Google search. It is beneficial to incorporate CodeMatcher into the Google search for practical usage.}

\section{Discussion}\label{disucss}
\subsection{Qualitative Analysis}\label{qualitative_analysis}

This subsection performs a qualitative analysis of CodeMatcher, DeepCS, CodeHow\bl{, and UNIF}. To compare these models, we classified all their returned methods into seven categories\bl{, namely whether the query can match the semantics of method name/body}. Table \ref{fig_classify} illustrates that the definitions and real query-code examples for each category. \bl{To be specific, the categories identified whether the method name or body of a searched code can reflect the semantics in query;} whether a code's method body is useless, i.e., an abstract method or a getter/setter; whether a method is a duplication of previously inspected one in the top-10 code list. Table \ref{tb_classify} lists the classification results for different models.

\begin{table}
    \centering
    \small
    \caption{Definitions of seven categories on search results and their query-code examples.}
    \begin{tabular}{|l|}
        \toprule
        \textbf{1. MM: Matched method name and Matched method body.}\\
        \bl{Query: create arraylist from array}\\
        public void \rd{createArrayListFromArray}()\{\\
        \quad{}\rd{String[] dogs = \{"Puppy", "Julie", "Tommy"\};}\\
        \quad{}\rd{List<String> doglist = Arrays.asList(dogs);}\\
        \quad{}assertEqual(3, dogsList.size());\\
        \}\\
        \midrule
        \textbf{2. NM: Not-matched method name but matched method body.}\\
        \bl{Query: how to declare an array}\\
        public void arrayCardinality(ParameterRegistration parameterRegistration)\{\\
        \quad{}\rd{Integer[] array = new Integer[]\{1, 2, 3\};}\\
        \quad{}Int arrayCardinality = procedures(parameterRegistration).arrayCardinality(array);\\
        \quad{}assertEquals(array.length, arrayCardinality);\\
        \}\\
        \midrule
        \textbf{3. MN: Matched method name but Not-matched method body.}\\
        \bl{Query: converting string to int in java}\\
        static Int \rd{convert}Status\rd{StringToInt}(String statusVal)\{\\
        \quad{}if (statusVal.equalsIgnoreCase(STATUS\_REGRESSION)) ||\\
        \quad{}\quad{}statusVal.equalsIgnoreCase(STATUS\_FAILED)\{\\
        \quad{}\quad{}return -1;\\
        \quad{}\} else if (statusVal.equalsIgnoreCase(STATUS\_PASSED))\{\\
        \quad{}\quad{}return 1;\\
        \quad{}\}\\
        \quad{}return 0;\\
        \}\\
        \midrule
        \textbf{4. MU: Matched method name but Useless method body.}\\
        \bl{Query: how can I initialise a static map}\\
        private void \rd{initialiseMap}(GoogleMap googleMap)\{\\
        \quad{}mMap = googleMap;\\
        \}\\
        \midrule
        \textbf{5. NU: Not-matched method name and Useless method body.}\\
        \bl{Query: converting iso 8601-compliant string to date}\\
        public static String convertDate2String(Date date)\{\\
        \quad{}return convertDate2String(date);\\
        \}\\
        \midrule
        \textbf{6. NN: Not-matched method name and Not-matched method body.}\\
        \bl{Query: how to read a large text file line by line using java}\\
        private String textLine(String name, long version, String value)\{\\
        \quad{}return String.format("name: \%s version: \%d value: \%s", name, version, value);\\
        \}\\
        \midrule
        \textbf{7. RM: Repeated Method.}\\
        \bl{Query: convert an inputstream to a string}\\
        public static String \rd{convertInputStreamToString}(InputStream inputStream)\{...\}\\
        private String \rd{convertInputStreamToString}(InputStream inputStream)\{...\}\\
        \bottomrule
    \end{tabular}
    \label{fig_classify}
\end{table}



\begin{table}[]
\small
\caption{Classification of \bl{1740} code search results into 7 categories for different models.}
\begin{tabular}{|c|l|rr|rrrrr|}
\toprule
\textbf{Queries}               & \textbf{Model} & \textbf{MM}  & \textbf{NM}  & \textbf{MN} & \textbf{MU} & \textbf{NU}  & \textbf{NN}   & \textbf{RM} \\ \midrule
\multirow{4}{*}{$Queries_{50}$}  & DeepCS         & 93 (18.6\%)  & 12 (2.4\%)   & 25 (5.0\%)  & 53 (10.6\%) & 97 (19.4\%)  & 218 (43.6\%)  & 2 (0.4\%)   \\ 
                               & CodeHow        & 34 (6.8\%)   & 73 (14.6\%)  & 2 (0.4\%)   & 1 (0.2\%)   & 4  (0.8\%)   & 331 (66.2\%)  & 55 (11.0\%) \\ 
                               & UNIF           & 60 (12.0\%)  & 23 (4.6\%)   & 0 (0.0\%)   & 7 (1.4\%)   & 18 (3.6\%)   & 392 (78.4\%)  & 0 (0.0\%)   \\ 
                               & CodeMatcher    & 285 (57.0\%) & 0 (0.0\%)    & 32 (6.4\%)  & 27 (5.4\%)  & 28 (5.6\%)   & 125 (25.0\%)  & 3 (0.6\%)   \\ \midrule
\multirow{4}{*}{$Queries_{99}$}  & DeepCS         & 59 (6.0\%)   & 48 (4.8\%)   & 0 (0.0\%)   & 3 (0.3\%)   & 69 (7.0\%)   & 811 (81.9\%)  & 0 (0.0\%)   \\ 
                               & CodeHow        & 31 (3.1\%)   & 155 (15.7\%) & 0 (0.0\%)   & 0 (0.0\%)   & 3 (0.3\%)    & 781 (78.9\%)  & 20 (2.0\%)  \\
                               & UNIF           & 118 (11.9\%) & 38 (3.8\%)   & 5 (0.5\%)   & 4 (0.4\%)   & 84 (8.5\%)   & 741 (74.8\%)  & 0 (0.0\%)   \\ 
                               & CodeMatcher    & 367 (37.1\%) & 0 (0.0\%)    & 52 (5.3\%)  & 89 (8.9\%)  & 0 (0.0\%)    & 473 (47.8\%)  & 9 (0.9\%)   \\ \midrule
\multirow{4}{*}{$Queries_{25}$}  & DeepCS         & 6 (2.4\%)    & 8 (3.2\%)    & 0 (0.0\%)   & 1 (0.4\%)   & 15 (6.0\%)   & 220 (8.0\%)   & 0 (0.0\%)   \\ 
                               & CodeHow        & 6 (2.4\%)    & 29 (11.6\%)  & 0 (0.0\%)   & 0 (0.0\%)   & 0 (0.0\%)    & 213 (85.2\%)  & 2 (0.8\%)   \\  
                               & UNIF           & 28 (11.2\%)  & 10 (4.0\%)   & 2 (0.8\%)   & 0 (0.0\%)   & 15 (6.0\%)   & 195 (78.0\%)  & 0 (0.0\%)   \\
                               & CodeMatcher    & 47 (18.8\%)  & 1 (0.4\%)    & 8 (3.2\%)   & 3 (1.2\%)   & 0 (0.0\%)    & 191 (76.4\%)  & 0 (0.0\%)   \\ \midrule
\multirow{4}{*}{$Queries_{all}$} & DeepCS         & 158 (9.1\%)  & 68 (3.9\%)   & 25 (1.4\%)  & 57 (3.3\%)  & 181 (10.4\%) & 1249 (71.8\%) & 2 (0.1\%)   \\ 
                               & CodeHow        & 71 (4.1\%)   & 257 (14.8\%) & 2 (0.1\%)   & 1 (0.1\%)   & 7 (0.4\%)    & 1325 (76.1\%) & 77 (4.4\%)  \\ 
                               & UNIF           & 206 (11.8\%) & 71 (4.1\%)   & 7 (0.4\%)   & 11 (0.6\%)  & 117 (6.7\%)  & 1328 (76.3\%) & 0 (0.0\%)   \\ 
                               & CodeMatcher    & 699 (40.2\%) & 1 (0.1\%)    & 92 (5.3\%)  & 119 (6.8\%) & 28 (1.6\%)   & 789 (45.3\%)  & 12 (0.7\%)  \\ \bottomrule
\end{tabular}
\label{tb_classify}
\end{table}

\subsubsection{The Reasons Why DeepCS \bl{and UNIF} Succeeded and Failed}

From the Table \ref{tb_classify}, we observed that the DeepCS\bl{/UNIF} obtained a \bl{13\%/15.9\%} success (MM and NM), where \bl{9.1\%/11.8\%} of success in MM was due to a correct semantic matching between query and method, as Table \ref{fig_classify}(1); the \bl{3.9\%/4.1\%} of success in NM implies that DeepCS\bl{/UNIF} can somewhat capture the semantics in code (i.e., API sequence and tokens) although the method name does not relate to the goal of a query as Table \ref{fig_classify}(2). However, there are \bl{87\%/88.2\%} of failed results, where \bl{0.1\%/0.0\%} of failures were caused by returning repeated methods (RM). The source code provided by Gu et al. \cite{gu2018deep} excluded the methods whose cosine similarity differences with related queries are larger than 0.01. But we observed that this judgment could not clear out repeated methods for \bl{the DL-based models} with some negligible difference, e.g., the modifier difference, as shown in Table \ref{fig_classify}(7). Meanwhile, \bl{1.4\%/0.4\%} of failures (MN) were caused by unmatched method body, because two methods for different usages may have the same method name, as exemplified in Table \ref{fig_classify}(3). 

Moreover, for the \bl{10.7\%/7.3\%} of failures (MU and NU), we found that DeepCS\bl{/UNIF} returned some useless methods that can be a getter/setter for a class, or contain abstract APIs with insufficient context to understand, such as the examples in Table \ref{fig_classify}(4-5). In this way, useless methods do not satisfy the requirement of the method-level code search since developers need to search and jump to related code. And the manual code jump will increase developers' code inspection time, and it is also uncertain how many jumps they need. Thus, the self-contained source code is advantageous for the method-level code search. In addition, for the most part \bl{(71.8\%/76.3\%)} of features (NN), DeepCS\bl{/UNIF} completely mismatched the code to queries, as illustrated in Table \ref{fig_classify}(6). We attribute these failures to the insufficient model training, because (1) DeepCS\bl{/UNIF} was optimized by pairs of method and Javadoc comment, but 500 epochs of training with randomly selected pairs cannot guarantee its sufficiency; (2) DeepCS\bl{/UNIF} assumed that the first line of Javadoc comment could well describe the goal of related code, but it is uncertain whether the used line is a satisfactory label or a noise; (3) during the model training, the optimization never stop because of the convergence of its loss function values.

\OB{1}{\bl{The DL-based models (DeepCS and UNIF)} can capture the semantics between queries and code methods. But its unsatisfactory performance is likely derived from the gap between the training data and our codebase.}

\subsubsection{CodeMatcher vs. DeepCS\bl{/UNIF}}

For the proposed model CodeMatcher that matches keywords in query with a method, Table \ref{tb_classify} shows that \bl{40.2\%} of code search succeeded due to the well-matched method name and body. However, there is no success from NM because wrongly combining keywords in the query only leads to unmatched code, i.e., MN \bl{(5.3\%)} and NN \bl{(45.3\%)}. This is because CodeMatcher cannot handle complex queries like DeepCS\bl{/UNIF} via the embedding technique. The \bl{8.4\%} of failures (MU and NU) on useless methods indicate that boosting the useful methods on a higher rank in terms of the percentage of JDK APIs is not the optimal solution, and directly removing those useless methods may be a better substitution. \bl{Moreover, we can observe that as CodeMatcher searched code methods based on how the tokens in methods matched the important query words sequentially, this sequential requirement would substantially exclude many duplicated methods for a query. After filtering out the redundant methods by simply comparing their MD5 hash values, the CodeMatcher can only return a limited number of repeated methods (RM=12).} \bl{Comparing with DeepCS/UNIF,} CodeMatcher returned more repeated methods. Thus, simply filtering redundant methods by their MD5 hash values, as described in Section \ref{method}, is not enough. A better choice would be comparing the API usages, data structure, and working flow in the method body. Comparing with DeepCS\bl{/UNIF}, the main advantage of CodeMatcher is the correct keywords matching between query and code, i.e., a high percentage of MM. However, CodeMatcher \bl{can hardly} handle partial matching only on the code body (i.e., \bl{NM=1}). But this is what DL-based models are good at (NM=\bl{68} for DeepCS \bl{and NM=71 for UNIF}) because they can capture high-level intent between query and method by joint embedding\bl{, although the NM are much smaller than MM for DeepCS/UNIF (MM=158 for DeepCS while MM=206 for UNIF).} Therefore, \bl{the advantages of DeepCS/UNIF} are good to compensate for CodeMatcher's disadvantages.


\OB{2}{CodeMatcher cannot address complex queries like \bl{DL-based models (DeepCS and UNIF)}. Meanwhile, CodeMatcher can accurately find code methods for simple queries and avoid the out-of-vocabulary issue in DeepCS \bl{and UNIF}.}

\subsubsection{CodeMatcher vs. CodeHow}

By analyzing the classification of these search results in Table \ref{tb_classify}, we can observe that CodeHow is good at matching a query with the method body (\bl{14.8\%} for NM), because it extends query with related official APIs so that method body can be well filtered in terms of internal APIs. However, the main reason for the failures is the unmatched keywords (\bl{76.1\%} for NN) since CodeHow ignores the importance of programming words and their sequence. The other main reason is that it does not exclude repeated methods (\bl{4.4\%} for RM). However, we can observe that CodeHow can compensate for the disadvantages of CodeMatcher, i.e., the difficulty in matching a query with the method body.

\OB{3}{The IR-based models CodeMatcher and CodeHow compensate with each other as they are good at matching a query with method name and body respectively.}

\subsection{\bl{Conciseness and Completeness}}\label{concise_complete}

\bl{Keivanloo et al. \cite{keivanloo2014spotting} indicated that, for a searched code, although the correctness (i.e., whether it is relevant to the search query) is important, developers also care about two other features. One feature is the conciseness, the ratio of irrelevant lines to the total lines. Lower conciseness indicates that a code contains no irrelevant lines. The other feature is the completeness, the number of addressed tasks divided by the total number of tasks, where the task includes the intent of the search query and other missed statements (well-typed, variable initialization, control flow, and exception handling) \cite{keivanloo2014spotting}. Besides, Keivanloo et al. \cite{keivanloo2014spotting} indicated that the code readability (whether the variable names are well-chosen) is also important. However, the readability is not easy to measure and the conciseness is the key component of readability (namely the searched code should use "as little code as possible" and show "the most basic version of the problem") \cite{buse2012synthesizing}. This is also the reason why Keivanloo et al. \cite{keivanloo2014spotting} only measured the conciseness and completeness of code.}

\begin{figure}
    \centering
    \includegraphics[width=0.8\columnwidth]{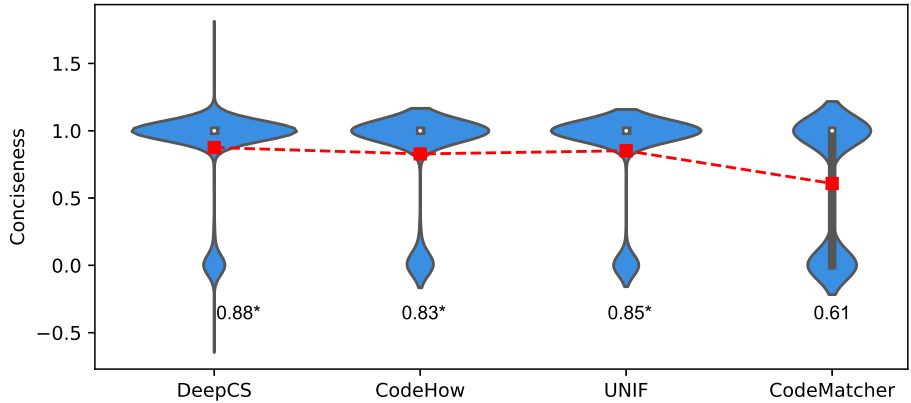}
    \caption{\bl{The conciseness of baseline models (DeepCS, CodeHow, UNIF, and CodeMatcher) for all the searched results; '*' indicates the significant difference (p-value $<$ 0.05) between a model and CodeMatcher, which is tested by the Wilcoxon signed-rank test \cite{wilcoxon1945individual}.}}
    \label{fig_concise}
\end{figure}

\begin{figure}
    \centering
    \includegraphics[width=0.8\columnwidth]{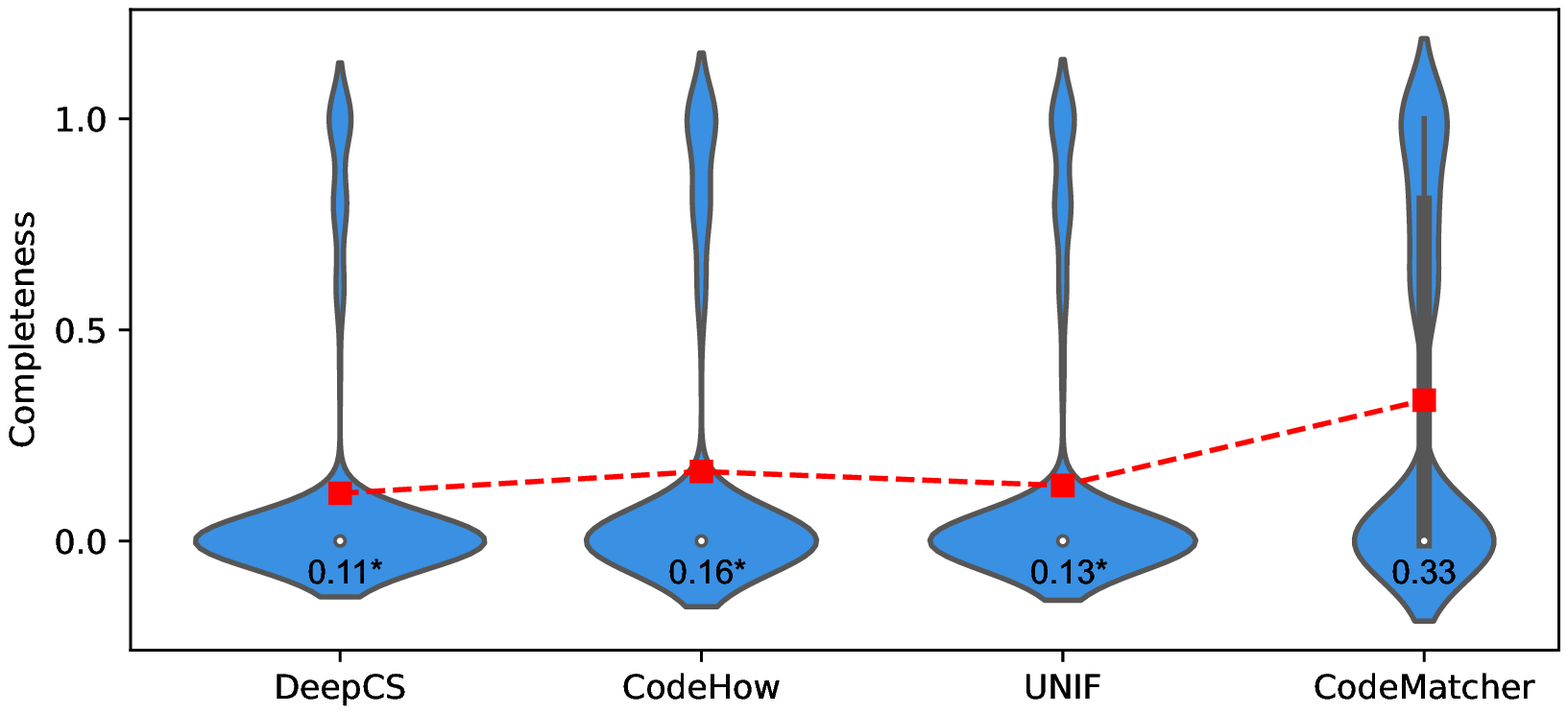}
    \caption{\bl{The completeness of baseline models (DeepCS, CodeHow, UNIF, and CodeMatcher) for all the searched results; '*' indicates the significant difference (p-value $<$ 0.05) between a model and CodeMatcher, which is tested by the Wilcoxon signed-rank test \cite{wilcoxon1945individual}.}}
    \label{fig_complete}
\end{figure}

\bl{Figs. \ref{fig_concise}-\ref{fig_complete} showed these two features of all the code searched by the CodeMatcher and baseline models (DeepCS, CodeHow, and UNIF). Fig. \ref{fig_concise} shows that the average conciseness of DeepCS, CodeHow, and UNIF are 0.88, 0.83, and 0.85 respectively. The proposed model CodeMatcher achieves a value of 0.61, outperforming the baselines by 30.7\%, 26.5\%, and 28.2\% respectively. After performing the Wilcoxon signed-rank test \cite{wilcoxon1945individual} between the results of CodeMatcher and a baseline model, the statistical result indicates that the advantages of CodeMatcher over baselines are significant (p-value<0.05). Moreover, Fig. \ref{fig_complete} shows that CodeMatcher also outperforms baseline models significantly in terms of the completeness with an average value of 0.33.}

\begin{figure}
    \centering
    \includegraphics[width=0.8\columnwidth]{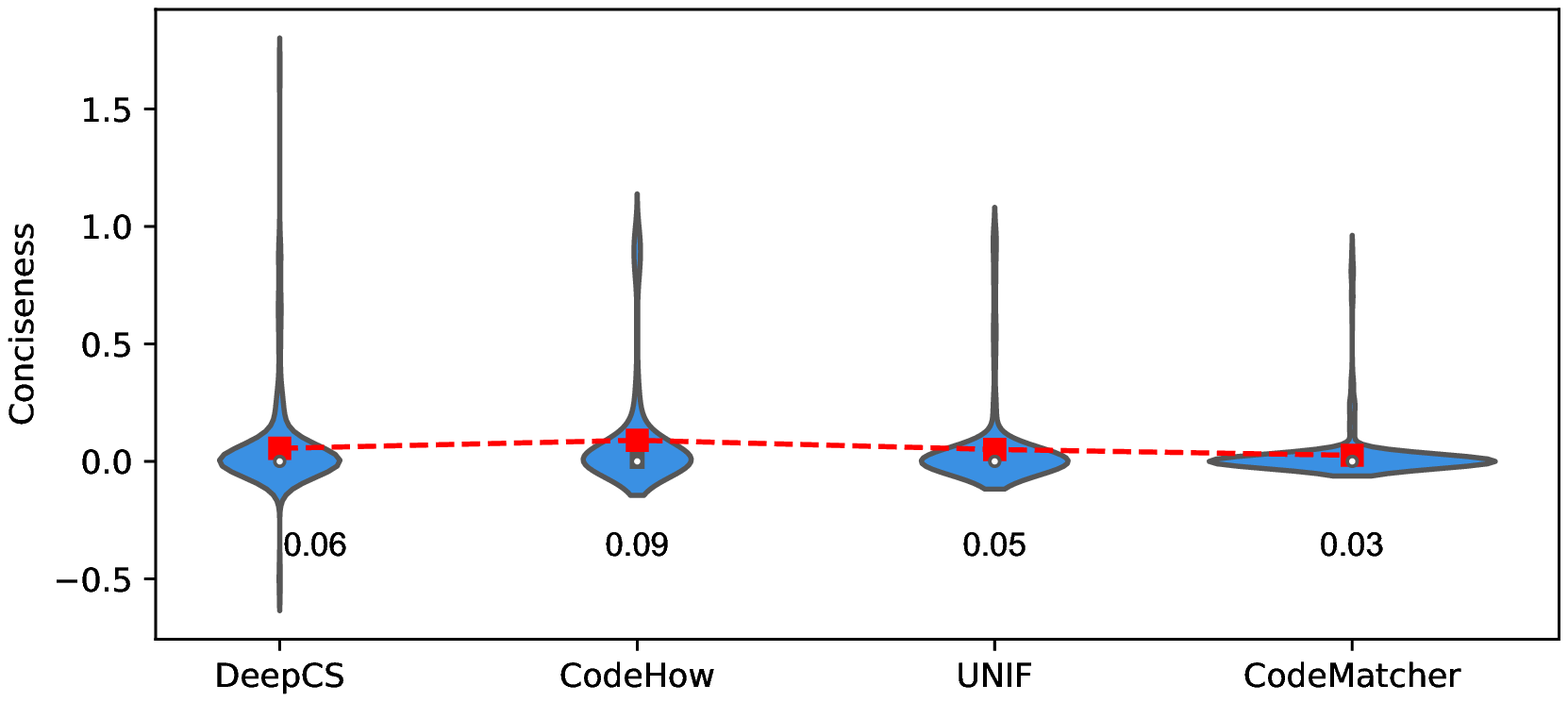}
    \caption{\bl{The conciseness of baseline models (DeepCS, CodeHow, UNIF, and CodeMatcher) for all the correctly searched results; no baseline showed significant difference (p-value $>$ 0.05) with CodeMatcher, which is tested by the two-sample Kolmogorov-Smirnov test \cite{massey1951kolmogorov} at a 5\% significance level.}}
    \label{fig_concise_short}
\end{figure}

\begin{figure}
    \centering
    \includegraphics[width=0.8\columnwidth]{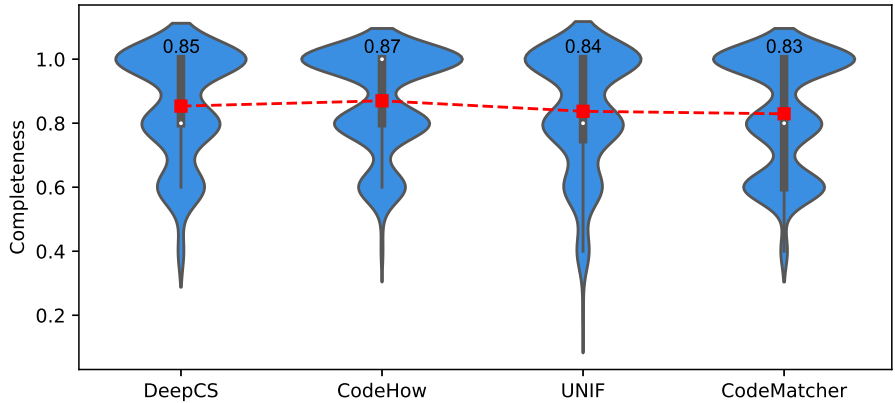}
    \caption{\bl{The completeness of baseline models (DeepCS, CodeHow, UNIF, and CodeMatcher) for all the correctly searched results; no baseline showed significant difference (p-value $>$ 0.05) with CodeMatcher, which is tested by the two-sample Kolmogorov-Smirnov test \cite{massey1951kolmogorov} at a 5\% significance level.}}
    \label{fig_complete_short}
\end{figure}

\bl{To investigate why CodeMatcher shows substantially better conciseness and completeness over baselines, we excluded the searched results irrelevant to the 174 queries for each model. As the left searched results are different for CodeMatcher and baselines, we applied the two-sample Kolmogorov-Smirnov test \cite{massey1951kolmogorov} at a 5\% significance level to estimate the statistical difference. Figs. \ref{fig_concise_short}-\ref{fig_complete_short} show that the CodeMatcher and baselines have no significant difference on conciseness and completeness for the relevant code. These results indicate that the number of returned code relevant to the query strongly affects the total conciseness and completeness. However, these results also implied that all the studied code search models do not consider the conciseness and completeness in the model. Therefore, we suggest that further studies take efforts to improve the quality of the searched code.}

\subsection{Why DeepCS \bl{and UNIF} Did Not Work Well on Our New Dataset?}\label{why}

One may notice that the performance of DeepCS \bl{and UNIF} are lower than the ones reported by Gu et al. \cite{gu2018deep} \bl{and Cambronero et al. \cite{cambronero2019deep} respectively}. To verify the validity of our re-ran DeepCS \bl{and re-implemented UNIF}, we tested our trained DeepCS \bl{and UNIF} not on our new testing data but the original testing data shared by Gu et al. \cite{gu2018deep}; we refer to the results of our experiment as DeepCS$_{old}$ and UNIF$_{old}$. Table \ref{tab_why} shows that the performance of DeepCS$_{old}$ is close to the one reported by Gu et al. \cite{gu2018deep} in terms of MRR (0.59 vs. 0.6)\bl{; the MRR of UNIF$_{old}$ is also close to the one reported by Cambronero et al. \cite{cambronero2019deep} (0.54 vs. 0.58)}. Thus, these results confirm the validity of the re-ran DeepCS model \bl{and our re-implemented UNIF model}. Moreover, not like \bl{Gu et al.'s} testing data, our testing data shares no overlap with \bl{their} training data. Therefore, the reduced performance on our testing data implies that the overlap affects the generalizability of DeepCS \bl{and UNIF}. 

Ideally, the performance of DeepCS \bl{and UNIF} can be improved by tuning with more training data. To reach the goal, we built the models DeepCS$_{tune}$ and UNIF$_{tune}$ tuned with more training data extracted from our codebase. Totally, we collected 1,283,445 commented code methods as training data instances following the training data extraction process described in \cite{gu2018deep}. In the model optimization step, we tuned the pre-trained DeepCS \bl{and UNIF} with the new training data in default settings, i.e., 500 epochs. Table \ref{tab_why} shows that DeepCS$_{tune}$ achieves a better performance with MRR \bl{0.38 (an improvement of 15.2\% over DeepCS) while the MRR of UNIF$_{tune}$ is 0.46 (an improvement of 12.2\% over UNIF)}. The experimental results imply that DeepCS \bl{and UNIF} did not work due to the gap between \bl{Gu et al.'s} training and our testing data. Tuning DeepCS \bl{and UNIF} with data generated from the testing data can mitigate the generalizability issue. However, their tuned performance (MRR = 0.38 \bl{for DeepCS and MRR = 0.46 for UNIF}) are still far from the result of CodeMatcher (MRR = \bl{0.60}).

We observed that this case is likely attributed to the limited training data and a fixed size of vocabulary. As shown in Table \ref{codebase}, our new testing data contains about 16 million methods in total but only 21.91\% of them have Javadoc comments. Moreover, only 7.73\% of the total methods can be used to tune DeepCS\bl{/UNIF} due to the limited vocabulary size. Therefore, DeepCS\bl{/UNIF} would have difficulty in comprehending the semantics of methods and their relationship to related Javadoc comments. To further analyze the impacts of the vocabulary, we investigate how it covers the words in the new testing data. Table \ref{tab:vocab} shows that more than 95\% of words in method components (API, name, and tokens) cannot be covered. In terms of the word frequency, there are 42.95\%, 6.22\%, and 12.75\% of new words that appeared in method API, name, and token respectively. Due to the out-of-vocabulary issue, more than half (52.91\%) of methods in testing data contain new words. Therefore, the trained DeepCS\bl{/UNIF} would have difficulty in understanding the semantics of methods in the new testing data.


\begin{table}[]
\caption{Performance comparison of DeepCS in different settings \bl{(old, testing model on DeepCS' original testing data \cite{gu2018deep}; tune, tuning model with data collected from our testing data and testing model on our testing data)}, where the model performance is measured by SuccessRate@1/5/10 (SR@1/5/10), Precision@1/5/10 (P@1/5/10), and MRR.}
\begin{tabular}{|c|l|ccc|ccc|c|}
\toprule
\textbf{Query}                 & \textbf{Model}  & \textbf{SR@1} & \textbf{SR@5} & \textbf{SR@10} & \textbf{P@1} & \textbf{P@5} & \textbf{P@10} & \textbf{MRR} \\ \midrule
\multirow{2}{*}{$Queries_{50}$}  & $DeepCS_{old}$  & 0.44          & 0.72          & 0.82           & 0.44         & 0.42         & 0.41          & 0.59         \\ 
                               & $UNIF_{old}$    & 0.38          & 0.72          & 0.80           & 0.38         & 0.39         & 0.33          & 0.54         \\ \midrule
\multirow{2}{*}{$Queries_{50}$}  & $DeepCS_{tune}$ & 0.30          & 0.62          & 0.72           & 0.30         & 0.31         & 0.30          & 0.44         \\ 
                               & $UNIF_{tune}$   & 0.26          & 0.56          & 0.64           & 0.26         & 0.26         & 0.19          & 0.41         \\ \midrule
\multirow{2}{*}{$Queries_{99}$}  & $DeepCS_{tune}$ & 0.24          & 0.42          & 0.47           & 0.24         & 0.20         & 0.15          & 0.36         \\ 
                               & $UNIF_{tune}$   & 0.33          & 0.64          & 0.71           & 0.33         & 0.32         & 0.26          & 0.49         \\ \midrule
\multirow{2}{*}{$Queries_{25}$}  & $DeepCS_{tune}$ & 0.16          & 0.36          & 0.40           & 0.16         & 0.18         & 0.10          & 0.30         \\
                               & $UNIF_{tune}$   & 0.36          & 0.52          & 0.52           & 0.36         & 0.22         & 0.17          & 0.48         \\ \midrule
\multirow{2}{*}{$Queries_{all}$} & $DeepCS_{tune}$ & 0.24          & 0.47          & 0.53           & 0.25         & 0.23         & 0.19          & 0.38         \\ 
                               & $UNIF_{tune}$   & 0.32          & 0.60          & 0.66           & 0.32         & 0.29         & 0.23          & 0.46         \\ \bottomrule
\end{tabular}
\label{tab_why}
\end{table}

\begin{table}
    \centering
    \caption{Statistics of words in our new testing data covered by DeepCS' \bl{and UNIF's} vocabulary, where \bl{DeepCS represents code by API, name, and token; meanwhile, UNIF represents code by name and token.}.}
    \begin{tabular}{|l|rrr|}
        \toprule
        \textbf{Item} & \textbf{Count} & \textbf{New} & \textbf{Percentage} \\
        \midrule
        Unique Words in API & 2,406,547 & 2,397,969& 99.64\%\\
        Unique Words in Name & 270,084 & 261,113 & 96.68\%\\ 
        Unique Words in Token & 1,610,147 & 1,602,435 & 99.52\%\\
        Words in API & 64,754,427 & 27,811,106 & 42.95\%\\
        Words in Name & 44,194,936 & 2,749,420 &  6.22\%\\
        Words in Token & 177,572,258 & 22,633,682 & 12.75\%\\
        Method & 16,611,025 & 8,788,967 & 52.91\%\\
        \bottomrule
    \end{tabular}
    \label{tab:vocab}
\end{table}

\subsection{\bl{Why is the Word Sequence Important?}}

\bl{To capture the sequential relationship between important words in query, the proposed model CodeMatcher considered it in the fuzzy search component and reranking strategies. However, one major question is that whether the sequence frequently occurred. To investigate this research question, we analyzed the studied 174 queries. We found that for 79 queries, the order of words that appear in the query plays an important role -- if the words in the query are reordered, the meaning will change. To analyze the root causes, we classified these 79 queries into four cases: \textit{1) the order of multiple tasks,} the query "read from file A and write to file B" is different from the one "write from file A and read to file" because exchanging the phrases "read from" and "write to" will change the intent of two tasks; \textit{2) data conversion,} "convert int to string" and "convert string to int" work differently; \textit{3) conditional job,} "sort map by values" contradicts with the semantics "sort values by map"; \textit{4) the exchanged core word,} "read property file" and "read file property" would be implemented in two ways because "property file" and "file property" are two completely different objects. The statistics in Table \ref{table_word_sequence} shows that among these 4 cases, the data conversion is the most frequent one.}

\begin{table}[]
\caption{\bl{Classification of cases with sequentially important query words.}}
\begin{tabular}{|c|l|l|c|r|}
\toprule
\textbf{No.}&\textbf{Case}   & \textbf{Description}                     & \textbf{Count} & \textbf{Percentage} \\ \midrule
1&The order of multiple tasks  & do A and B                               & 9              & 11.4\%              \\
2&Data conversion & do (from) A to B                         & 38             & 48.1\%             \\
3&Conditional job & do A with/via/in/over/based/by/when/if B & 17             & 21.5\%              \\
4&The exchanged core word        & do A B                                   & 15             & 19.0\%              \\
-&\textbf{Total}  & -                                        & 79             & 100.0\%             \\ \bottomrule
\end{tabular}
\label{table_word_sequence}
\end{table}

\section{Implication}\label{implication}


\subsection{Pros and Cons of DL-based Model and IR-Based Model}\label{pros}

The DL-based model has three major advantages over the IR-based model. One is language processing ability. By leveraging the embedding technique, it can better address complex queries and tolerate errors to some extent according to Section \ref{disucss}. The second is the bilingual learning ability. With the joint embedding framework between a query written in natural language and a code implemented in programming language, their mapping relationship can be well learned as described in Section \ref{rq1}. At last, the DL-based model may require less upfront cost than IR-based model, because the former requires much less domain knowledge and feature engineering like the IR-based model. However, \bl{the DL-based models (DeepCS and UNIF) have} a limitation on a new and dynamic codebase, because \bl{the studied models} suffer from overfitting and out-of-vocabulary issues as discussed in Section \ref{why}. But these problems do not occur for the IR-based models (CodeHow and CodeMatcher). Meanwhile, running the IR-based model is substantially more efficient over the DL-based model, because the IR-based model requires no model training and can speed up the code search process using a framework like Elasticsearch, as shown in Section \ref{codematcher_evaluation}. To sum up, the two kinds of studied code search models complement each other, therefore it is suggested to balance their pros and cons and make a fusion in the future. 

\SG{1}{Combine the advantages of IR-based and DL-based models.}

\subsection{The Importance of Method Name}

Section \ref{why_well} indicates that CodeMatcher works well mainly because it can precisely match the \bl{semantics between query and the method name}, where CodeMatcher assigns higher importance on programming words (e.g., Inputstream or String) and considers their sequence in the query. Besides, Section \ref{disucss} also shows that if a method precisely matched \bl{the semantics} in query such method is likely to contain expected implementation in the method body. This is because the method name is very similar to the query: \textit{(1) writing in natural language.} There is \bl{less} semantic gap between query and method name; \textit{(2) short in text.} They usually use the same keywords in order; \textit{(3) specific to code implementation.} Their semantic relationship to code implementation is usually the same and straightforward. Therefore, a code search engine should assign higher weights on method names, no matter for the DL-based or IR-based model.

Moreover, although CodeMatcher is capable of handling synonyms in a query by using the WordNet as described in Section \ref{method}, it has three limitations: \textit{(1) abbreviation.} It cannot match the word 'initialize' in a query to the method named with 'ini'; \textit{(2) Acronym.} The method named with 'RMSD' should be missed for a query with the keyword "root mean square deviation"; \textit{(3) low quality of method naming.} The method name is not a correct abstraction on its code implementation. Meanwhile, other code search models (e.g., DeepCS and CodeHow) also do not consider these situations. One solution to these limitations is to increase the scale of the codebase, because the increased search space may include the methods whose names are similar to search queries (e.g, using abbreviations or acronyms). However, to solve these challenges, maybe the best way is to require developers strictly following a method naming standard in the beginning. For example, a developer follows the Google Java style guide\footnote{https://google.github.io/styleguide/javaguide.html\#s5-naming} and writes method name in verb (phases) with commonly used words, instead of self-defined synonym, acronym, and abbreviation. As to the existing large-scale source code in GitHub, maybe a better solution is to format their method names in a standard and unified way.

\SG{2}{Method name has a significant role in code search; improving the quality of developers' method names helps code search.}

\section{Threats to Validation and Model Limitation}\label{threat}

There are some threats affecting the validity of our experimental results and conclusions. 

\vspace{5pt}\noindent\textbf{Manual Evaluation.} 
The relevancy of returned code methods to the studied queries was manually identified, which could suffer from subjectivity bias. To mitigate this threat, the manual analysis was performed independently by two developers from Baidu inc. \bl{For the first query set ($Queries_{50}$),} they reach a substantial agreement in terms of the value (0.62) of Cohen's Kappa static \cite{viera2005understanding}; and if developers disagreed on a relevancy identification, they performed an open discussion to resolve it. \bl{For the second and third query sets ($Queries_{99}$ and $Queries_{25}$), the agreement was improved to 0.72 in terms of the Cohen's Kappa static due to developers' increased evaluation experiences.} In the future, we will mitigate this threat by inviting more developers. Moreover, in the relevancy identification, we only consider the top-10 returned code results following Gu et al. \cite{gu2018deep}. However, in the real-world code search, this setting is reasonable because developers would like to inspect the top-10 results and ignore the remaining due to the impacts of developers' time and patience.

\vspace{5pt}\noindent\textbf{Limited Queries and Java Codebase.} 
Following Gu et al. \cite{gu2018deep}, we evaluated the model with popular questions from Stack Overflow, which may not be representative of all possible queries for code search engines. To mitigate this threat, the selected top-50 queries are the most frequently asked questions collected by Gu et al. \cite{gu2018deep} in a systematic procedure, as referred to in Section \ref{exp}. \bl{We also extended 99 more queries that provided in the CodeSearchNet challenge \cite{husain2019codesearchnet} and 25 more queries from two related studies \cite{mishne2012typestate,keivanloo2014spotting}.} \bl{Besides, many studied queries are highly related to popular APIs, so that the CodeMatcher may not work for queries with less popular APIs. In this case, the performance of CodeMatcher could be overestimated.} In the future, we will extend the \bl{scale, scope, and variety} of the code search queries. \bl{We also plan to investigate how to automatically evaluate model performance on a large-scale codebase}. Furthermore, we performed the experiments with large-scale open-source Java repositories. But we have not evaluated repositories in other programming languages, though the idea of extending CodeMatcher to any language is easy and applicable. Moreover, we collected about 41k GitHub projects with high-quality code (i.e., more than 5 stars) as the codebase. But such search space is likely to overestimate a model, because such projects are going to have more accurate Javadoc and generally cleaner, easier to understand code that is more likely to be commented. Although we extended Gu et al.'s \cite{gu2018deep} codebase (around 1k projects with at least 20 stars) on a larger scale, this situation cannot be ignored. We plan to extend our codebase more in the near future.  

\vspace{5pt}\noindent\textbf{Baseline Reproduction.}
To estimate DeepCS on our testing data, we preprocessed the testing data according to Gu et al. \cite{gu2018deep} although the source code and training data have no difference. Meanwhile, we re-implemented the baseline CodeHow \bl{and UNIF strictly following the paper \cite{lv2015codehow} and \cite{cambronero2019deep} respectively} because their authors provide no source code. Our baseline reproductions may threaten the validity of our model. To mitigate this threat, we double-checked our code and also present all the replication packages in public as described in Section \ref{baselines}.

\vspace{5pt}\noindent\textbf{Model Limitation.}
Like all the existing code search models, the proposed solution CodeMatcher may not work if a search query shares no irrelevant words with the methods in codebase. To address this limitation, CodeMatcher replaced the words that do not appeared in a codebase by their synonyms extracted from the codebase. This limitation could be further mitigated by combining CodeMatcher with a DL-based model as described in Section \ref{pros}. We intend to investigate this research in the near future. \bl{Section \ref{rq4} indicated the Stanford Parser used in CodeMatcher is not accurate. Although the corrected results only improve the model performance slightly for our studies, we cannot ignore this limitation in the further studies.}

\section{Related Work}\label{related}

\vspace{5pt}\noindent\textbf{Categories of Code Search.}
In the software development, developers may directly search existing applications to work on \cite{bajracharya2012analyzing}, and many application search engines have been built \cite{grechanik2010search,robillard2009recommendation,liu2018recommending,mcmillan2012detecting,zhang2017detecting,liu2018poster}. However, application-level search is not frequently used during the development, and more than 90\% of developers' search efforts are used for searching code snippets (e.g., code method) \cite{bajracharya2012analyzing}. For this reason, method-level code search has been studied for decades \cite{reiss2009semantics,lucredio2004survey,mili1998survey,mili1995reusing}, and this study follows this type of code search.

The research objective of method-level code search is to investigate the mechanics of developers' searching behaviors \cite{sadowski2015developers,xia2017developers,rahman2018evaluating,bajracharya2012analyzing,li2013help,keivanloo2014spotting}, and build a model to fill the semantic gap between natural language (i.e., search query) and programming language (i.e., method source code) \cite{gu2018deep,lv2015codehow,bajracharya2006sourcerer,lu2015query,li2016relationship}. Better code search techniques can boost the rapid software development \cite{mcmillan2012recommending} and promote other search-based researches, such as program synthesis \cite{galenson2014codehint,gvero2015interactive}, code completion \cite{raychev2014code,nguyen2012graph,robbes2010improving}, program repair \cite{le2016history,ke2015repairing}, and mining software knowledge \cite{gabel2010study}.

For the setting of method-level code search, this study works on the 'query-codebase' search following previous studies \cite{gu2018deep,lv2015codehow}. The query in natural language is commonly used as developers' search input, although method declarations and test cases can complement and clarify developers' specification \cite{reiss2009semantics,lemos2007codegenie,lemos2011test}. But we cannot assume that developers would always provide this information. Moreover, the codebase like GitHub is usually used for code retrieval. Because the stored code are large-scale and ready-to-use \cite{gu2018deep,lv2015codehow,bajracharya2006sourcerer}, although there are some useful code provided in Stack Overflow \cite{campbell2017nlp2code} or software development tutorials \cite{ponzanelli2016too}. 

\vspace{5pt}\noindent\textbf{Evolution of Method-Level Code Search.}
At the beginning of this study, researchers just regarded code as plain texts, and simply applied the capabilities of web search engines into code search\cite{krugler2013krugle}. Google Code Search \cite{kim2010towards}, Koders \cite{bajracharya2012analyzing} and Krugle \cite{krugler2013krugle} were few promising systems \cite{bajracharya2006sourcerer}. Later, many researchers attributed the challenge of code search to the understanding of query in natural language. To generate correct keywords for a search engine, many existing works focused on query expansion and reformulation \cite{hill2011improving,haiduc2013automatic,lu2015query,mcmillan2011portfolio,ponzanelli2014mining}. For example, Hill et al. \cite{hill2011improving} rephrased queries according to the context and semantic role of query words within the method signature. Haiduc et al. \cite{haiduc2013automatic} proposed Refoqus that reformulated query by choice of reformulation strategies, and a machine learning model recommends which strategy to use based on query properties. Lu et al. \cite{lu2015query} extended a query with synonyms generated from WordNet \cite{leacock1998combining}. McMillan et al. \cite{mcmillan2011portfolio} proposed Portfolio, which returns a chain of functions through keyword matching and speeds up search by PageRank \cite{page1999pagerank}. All query processing models can outperform early search engines, such as the substantial improvements of Portfolio over Google Code Search and Koders \cite{mcmillan2011portfolio}.

However, source code is more than just a plain text, and it contains abundant programming knowledge. Thus, matching code text with keywords is far from enough \cite{bajracharya2006sourcerer}. To push the code search study forward,
Bajracharya et al. \cite{bajracharya2006sourcerer} proposed Sourcerer, an IR-based code search engine that combines the textual content of a program with structural information. Moreover, researchers observed that the API usage is a key to understand code and its specification. Following this intuition, Li et al. \cite{li2016relationship} proposed RACS, a search framework for JavaScript that considers API relationships, including their sequencing, condition, callback relationships, etc. Meanwhile, Lv et al. \cite{lv2015codehow} proposed CodeHow, a code search engine that confers related APIs from a query and matches them with code by an extended Boolean model. Its validity and usefulness were validated by Microsoft developers. Similarly, Feng et al. \cite{zhang2017expanding} proposed a model to expand query with semantically related API class names and search the best-matched source code.

Recently, Gu et al. \cite{gu2018deep} proposed a deep-learning-based model DeepCS, which jointly embeds method in programming language and query in natural languages, and search methods by comparing the similarity between query and candidate methods. Their experiments show that DeepCS can significantly outperform two representative models, Sourcerer \cite{bajracharya2006sourcerer} and CodeHow \cite{lv2015codehow}. 
They attributed these improvements to the successful application of a deep learning model, which considers both code property (e.g., API sequence, programming tokens) and text property (e.g., token synonymous, rephrasing, sequencing) in a unified model. \bl{Recently, researchers also provided some improved DL-based models. Cambronero et al. \cite{cambronero2019deep} developed the model UNIF with much simpler network than DeepCS. Generally, UNIF embeds code and query with embedding layers pre-trained by the fastText \cite{bojanowski2017enriching}, and the embedded code and query were attached with attention layer and average layer respectively. Wan et al. \cite{wan2019multi} provided a model called MMAN (Multi-Model Attention Network) with complex network. Different from DeepCS, MMAN embeds code by a fusion of three neural networks: one LSTM for the sequential tokens
of code, a Tree-LSTM for the AST of code, and a GGNN (Gated
Graph Neural Network) for the control flow graph of code. Feng et al. \cite{feng2020codebert} fine-tuned the heavy-weight DL model BERT \cite{devlin2018bert} on the code search task with pairs of natural language text and code.} As described in Section \ref{method}, our proposed CodeMatcher can be regarded as a simplified model of DeepCS that also considers code and text property in search but implemented in the way of keyword matching. Experimental results showed that CodeMatcher substantially outperforms DeepCS \bl{and UNIF} in searching accuracy and time efficiency. Also, CodeMatcher shows better performance over the IR-based model CodeHow.

\section{conclusion}\label{conclude}

The challenge for code search is how to fill the semantic gap between a query written in natural language and a code implemented in programming language. Recently, Gu et al. \cite{gu2018deep} proposed a DL-based model DeepCS that leverages the DL technique to jointly embed query and method code into a shared high-dimensional vector space, where methods related to a query are retrieved by their vector similarities. However, the working process of DeepCS is complicated and time-consuming. In this paper, we proposed a simple and faster model named CodeMatcher, which leverages the IR technique to simplify DeepCS but inherits the advantageous features in DeepCS.

To verify the model validity, we collected a large scale codebase from GitHub to compare the performance of CodeMatcher and \bl{baseline models (DeepCS, CodeHow, and UNIF)}. Experimental results show that CodeMatcher (MRR = \bl{0.60}) substantially outperforms \bl{DeepCS, CodeHow, and UNIF by 82\%, 62\%, and 46\%} respectively in terms of MRR. On the time-efficiency, CodeMatcher is over \bl{1.2k and 1.5k} times faster than the DL-based models DeepCS and UNIF. Besides, it works \bl{8 times faster than} the IR-based model CodeHow. The above experimental results indicate that the CodeMatcher simplified from DeepCS is reasonable and valuable. Besides, we also found that CodeMatcher works well because it can correctly map the semantics of queries to code methods.

Moreover, we also compare CodeMatcher with two existing online code search engines, GitHub and Google search. Experimental results showed that CodeMatcher outperforms these two search engines by \bl{46\% and 33\%} respectively in terms of MRR. Comparing with Google search, the major advantage of CodeMatcher is to correctly recommend code method for query in the first place. Thus, the experimental result shows when we incorporated CodeMatcher to Google search, the MRR can be further improved to \bl{0.64}. These results imply the merit of CodeMatcher for practical usage.

Finally, we conducted an in-depth qualitative analysis on results of DL-based and IR-based models, and provided some suggestions for code search: it is necessary to combine the advantages of the IR-based and DL-based models together in the future; method name plays a significant role in code search because it is written in natural language as queries, and improving the quality of method names will help code search.



\bibliographystyle{ACM-Reference-Format}
\bibliography{reference}

\end{document}